\pdfoutput=1

\documentclass[%
 reprint,
 superscriptaddress,
 nofootinbib,
 amsmath,amssymb,
 aps,
 showkeys,
 prb,
]{revtex4-2}

\renewcommand{\selectlanguage}[1]{} 
\usepackage[dvipsnames]{xcolor} 
\usepackage{graphicx} 
\usepackage{dcolumn} 
\usepackage{bm} 
\usepackage[separate-uncertainty=true]{siunitx} 
\usepackage{physics2}
\usepackage{hyperref} 
\usepackage[capitalise]{cleveref} 
\usepackage{orcidlink}
\usepackage[version=4]{mhchem}
\usepackage{amsthm}
\usepackage{cancel}
\usepackage{multirow}
\usepackage[caption=false]{subfig}

\usepackage{listings} 
\usepackage{accsupp} 

\usepackage{pdfpages}
\usepackage{pgffor}
\makeatletter
\AtBeginDocument{\let\LS@rot\@undefined}
\makeatother

\graphicspath{{Figures/}}
\usephysicsmodule{ab,ab.legacy} 

\definecolor{Code}{rgb}{0,0,0} 
\definecolor{Decorators}{rgb}{0.5,0.5,0.5} 
\definecolor{Numbers}{rgb}{0.5,0,0} 
\definecolor{MatchingBrackets}{rgb}{0.25,0.5,0.5} 
\definecolor{Keywords}{rgb}{0,0,1} 
\definecolor{self}{rgb}{0,0,0} 
\definecolor{Strings}{rgb}{0,0.63,0} 
\definecolor{Comments}{rgb}{0,0.63,0} 
\definecolor{Backquotes}{rgb}{0,0,0} 
\definecolor{Classname}{rgb}{0,0,0} 
\definecolor{FunctionName}{rgb}{0,0,0} 
\definecolor{Operators}{rgb}{0,0,0} 
\definecolor{Background}{rgb}{0.98,0.98,0.98} 

\newcommand{\emptyaccsupp}[1]{\BeginAccSupp{method=escape,ActualText={}}#1\EndAccSupp{}}
\newcommand{\ccite}[1]{Ref.~\cite{#1}}
\newcommand{\crefSubFigRef}[2]{\crefformat{figure}{Fig.~##2##1{#2}##3}%
  \cref{#1}\crefformat{figure}{Fig.~##2##1##3}}
\makeatletter
\newcommand{\vb}[1]{\boldsymbol{\mathbf{#1}}}
\newcommand\va[1]{\@ifstar{\vec{#1}}{\vec{\mathrm{#1}}}}
\newcommand\vu[1]{\@ifstar{\hat{\boldsymbol{#1}}}{\hat{\mathbf{#1}}}}
\makeatother
\newcommand\bcdot{\boldsymbol{\cdot}}
\newcommand{\code}[1]{\texttt{#1}}
\newcommand{\python}[1]{\lstinline{#1}}

\newcommand{\numberthis}{\addtocounter{equation}{1}\tag{\theequation}}
\newcommand{\mumax}[1][3]{$\mathsf{mumax}^{#1}$} 
\newcommand{\spinengine}{SpinENGINE} 
\newcommand{\hotspice}{Hotspice}
\newcommand{\kB}{\ifmmode k_\mathrm{B} \else $k_\mathrm{B}$ \fi}
\newcommand{\kBT}{\ifmmode k_\mathrm{B}T \else $k_\mathrm{B}T$ \fi}

\begin{document}

\title{The design, verification, and applications of Hotspice:\\ a Monte Carlo simulator for artificial spin ice}

\def\affilDynamat{DyNaMat, Department of Solid State Sciences, Ghent University, 9000 Ghent, Belgium}
\def\affilETH{Laboratory for Mesoscopic Systems, Department of Materials, ETH Zurich, 8093 Zurich, Switzerland}
\def\affilPSI{Laboratory for Multiscale Materials Experiments, PSI Center for Neutron and Muon Sciences, Forschungsstrasse 111, 5232 Villigen PSI, Switzerland}
\author{Jonathan Maes\orcidlink{0000-0003-4973-8383}}\affiliation{\affilDynamat}
\author{Diego De Gusem\orcidlink{0009-0008-3341-0555}}\affiliation{\affilDynamat}
\author{Ian Lateur\orcidlink{0009-0003-8283-4083}}\affiliation{\affilDynamat}
\author{Jonathan Leliaert\orcidlink{0000-0001-8778-3092}}\affiliation{\affilDynamat}
\author{Aleksandr Kurenkov\orcidlink{0000-0002-0918-9554}}\affiliation{\affilETH}\affiliation{\affilPSI}
\author{Bartel Van Waeyenberge\orcidlink{0000-0001-7523-1661}}\affiliation{\affilDynamat}

\date{\today}

\begin{abstract}
    We present \hotspice{}, a Monte Carlo simulation software designed to capture the dynamics and equilibrium states of Artificial Spin Ice (ASI) systems with both in-plane (IP) and out-of-plane (OOP) geometries. An Ising-like model is used where each nanomagnet is represented as a macrospin, with switching events driven by thermal fluctuations, magnetostatic interactions, and external fields. To improve simulation accuracy, we explore the impact of several corrections to this model, concerning for example the calculation of the dipole interaction in IP and OOP ASI, as well as the impact of allowing asymmetric rather than symmetric energy barriers between stable states. We validate these enhancements by comparing simulation results with experimental data for pinwheel and kagome ASI lattices, demonstrating how these corrections enable a more accurate simulation of the behavior of these systems. We finish with a demonstration of `clocking' in pinwheel and OOP square ASI as an example of reservoir computing.
\end{abstract}

\maketitle

\section{Introduction}
\textit{Artificial Spin Ice} (ASI) are arrays of magnetostatically coupled nanomagnets arranged on a lattice.~\cite{RC_ASI, flatspin}
These nanomagnets are made sufficiently small -- usually a few tens of nanometers thick with lateral dimensions of the order of \qty{100}{\nano\meter}~\cite{nisoli2013colloquium} -- such that it is energetically favorable for them to have an almost homogeneous magnetization.~\cite{BrownThermalFluctuations,neel1949theorie,Kittel_TheoryFMDomains,Dipolar2Dparticles} 
Nanomagnets are designed to exhibit uniaxial anisotropy, meaning their magnetic moment prefers to align along the so-called ``easy axis''.~\cite{nisoli2013colloquium} Their typically flat geometry causes the magnetization to prefer an in-plane orientation.~\cite{neel1949theorie} 
Out-of-plane (OOP) magnets can be realized by using interfacial anisotropy to counteract this shape anisotropy, while for in-plane (IP) magnets a preferential axis is typically created by giving them an elongated shape: the easy axis then aligns with the long axis of the magnet. \par
Thus, each nanomagnet behaves as a bistable macrospin, with the magnetic moment pointing in either direction along the easy axis. Transitions between these stable states can occur spontaneously at elevated temperatures if the energy barrier posed by shape anisotropy is not too high.~\cite{BrownThermalFluctuations,CoerciveFieldReversal} Switching can also be facilitated by a magnetic field (of external origin or originating from neighboring magnets), or by more intricate methods such as current-induced torques.~\cite{SOT_FM_AFM,brataas2012current} \par
ASI systems were originally inspired by the magnetic frustration -- the presence of competing interactions which cannot simultaneously all be satisfied -- observed in natural spin ices found in certain crystal structures.
Spin ices, in turn, got their name from water ice, which is the prototypical system exhibiting geometrical frustration.~\cite{nisoli2013colloquium,ZeroPointEntropy,heyderman2013artificial,MagnetizationDynamicsASI}
Frustration can lead to correlations and collective behavior that the individual elements would not exhibit by themselves, resulting in nontrivial dynamics and complex magnetic orderings.~\cite{AdvancesASI,ASI_computation,ApparentFMpinwheel} \par
ASI is very attractive in comparison to its natural spin ice counterpart. Modern nanofabrication methods such as electron beam lithography allow any arbitrary geometry to be realized and many other aspects of the system to be engineered, offering enormous freedom to the designer.~\cite{AdvancesASI,ASI_computation}
Furthermore, the mesoscopic size of ASI (a few hundred nanometers) enables direct observation of their magnetic degrees of freedom through a variety of microscopy techniques, in contrast to natural spin ices where imaging individual spins is not feasible.~\cite{nisoli2013colloquium,freeman2001advances}
Hence, the controlled, tunable and easily measurable environment offered by ASI has enabled the study of complex phenomena such as phase transitions with ordered domains~\cite{ApparentFMpinwheel,sklenar2019field,MeltingASI,ImagingBridgedKagome,sendetskyi2019continuous,lou2023competing,branford2012emerging} or glasslike behavior~\cite{wang2006artificial,ZeroPointEntropy}, vertex-based frustration~\cite{morrison2013unhappy} leading to emergent topological structures such as monopoles and Dirac strings~\cite{ObservationMonopoleASI,mengotti2011kagome}, chiral dynamics...~\cite{branford2012emerging,EmergentChiralityRatchet}
ASI also holds significant potential for computational applications~\cite{heyderman2022spin}, both in conventional logic~\cite{ComputationalLogic_2018,gypens2018balanced,EngineeringRelaxationComputation} and neuromorphic computing.~\cite{ASI_computation,RC_RecentAdvances}

\par

The dynamics of ASI can be simulated using micromagnetic codes, such as the finite-difference-based \mumax[3]~\cite{MuMax3} and OOMMF~\cite{OOMMF} or the finite-element-based Nmag~\cite{Nmag}, which capture the magnetization dynamics of individual nanomagnets in great detail. 
However, the time between successive switches of a nanomagnet is not necessarily similar to the timescale of micromagnetics: when the simulated time extends beyond several microseconds, simulating even a modest number of magnets -- on the order of several dozen -- becomes computationally unfeasible.~\cite{leo2021chiral} \par
To address these limitations, specialized ASI simulation tools have been developed, such as the flatspin simulator~\cite{flatspin}, which implements deterministic spin flipping via a Stoner-Wohlfarth model.~\cite{StonerWohlfarth2008} Using such higher-level approximations enables the study of collective behavior in much larger systems and over far longer timescales than is feasible with micromagnetic codes, though at the cost that the internal magnetization structure of individual nanomagnets is no longer simulated in detail. Additionally, Monte Carlo methods are often used to simulate both IP~\cite{Qi2008,Cugliandolo2017,LocalizedFrustratedKagome,Brunn2021,Farhan2013,ApparentFMpinwheel} 
and OOP~\cite{Chioar2014,PerpendicularMagnetizationASI} 
ASI. However, these are typically specialized to a select few lattice geometries.
Early works often accounted only for nearest-neighbor (NN) interactions, whose strength was often arbitrarily set or calculated separately using micromagnetic codes.~\cite{Qi2008,PerpendicularMagnetizationASI} 
More recently, the importance of long-range magnetostatic interactions has been emphasized to improve correspondence to experiments.~\cite{Chioar2014,Rougemaille2011,Brunn2021}
As such, vertex~\cite{gilbert2014emergent,Saglam2022Tetris,Goryca2021Plasma,MeltingASI} or hybrid NN/vertex charge~\cite{Canals2016,zhang2013crystallites} models have become more prevalent, especially for square-lattice ASI. 
Beyond-NN interaction models, including full magnetostatically coupled models~\cite{ApparentFMpinwheel,mengotti2011kagome}, are also used for e.g. square-~\cite{Brunn2021,Farhan2013,sklenar2019field} and triangular-based~\cite{Chioar2014,Rougemaille2011,Hofhuis2020} lattices. \par 
Our goal was to blend these two approaches, resulting in \hotspice{}: a versatile Monte Carlo simulator meant to capture ASI physics with minimal arbitrary parameters, allowing various lattice configurations to be evaluated. This software approximates each single-domain nanomagnet as a single Ising spin, associating energies with its various states. The importance of considering long-range magnetostatic interactions, particularly for OOP systems, has been highlighted by Chioar~\textit{et al.}~\cite{Chioar2014}, so \hotspice{} explicitly accounts for the magnetostatic interaction between all magnets. Throughout this paper, we investigate several model variants to assess their accuracy in simulating the behavior of ASI. These variants differ in their calculation of the magnetostatic interactions, the use of symmetric versus asymmetric energy barriers, and their choice of update algorithms.

\section{Design}
\hotspice{} is a Monte Carlo simulation software implemented as a Python package, designed to model both in-plane (IP) and out-of-plane (OOP) ASI systems containing thousands of magnets over arbitrary timescales. It was developed to explore reservoir computing in ASI, as discussed in~\cref{sec:clocking}. Simulations can be performed on either CPU or GPU, with the optimal hardware choice depending on the size of the ASI and the update scheme used. Both open and periodic boundary conditions (PBC) are possible. \par
In this section, we describe the model implemented in \hotspice{} and discuss several variants of this model. 

\subsection{Model}
Because the magnetization prefers to align along the fixed easy axis of a nanomagnet, it is natural to use an Ising-like approximation where the position $\vb{r}_i$, axis $\vb{u}_i$, and size of magnetic moment\footnote{The size of the magnetic moment $\mu_i$ corresponds to the total ground state magnetic moment $\abs{\int_{\Omega_i} \vb{M}(\vb{r})d\vb{r}}$, with $\Omega_i$ the shape of magnet $i$ and $\vb{M}(\vb{r})$ its magnetization in the twofold degenerate ground state. Due to edge relaxation effects, this value is slightly smaller than $M_\mathrm{sat} V_i$.} $\mu_i$ of each magnet are fixed, allowing the magnets to only switch between the `up' and `down' states. Thus, the total magnetic moment vector of magnet $i$ can be expressed as $\vb{\mu}_i = s_i \mu_i \vb{u}_i$, where $s_i = \pm 1$ and $\abs{\vb{u}_i} = 1$. \par 
The switching rate between these states is determined by the temperature $T$ and the effective energy barrier $\widetilde{E_\mathrm{B}}$ separating them. For an isolated nanomagnet, the energy barrier $E_\mathrm{B} = K_\mathrm{u} V$\footnote{This is valid for switching by coherent rotation. Similar to the calculation of $\mu$, edge relaxation effects may cause the effective volume to be slightly smaller.} originates from its uniaxial shape anisotropy $K_\mathrm{u}$. $E_\mathrm{B}$ can be estimated from e.g. micromagnetic simulations. Interactions with other magnets or external fields modify the energy landscape, leading to an effective barrier $\widetilde{E_\mathrm{B}}$.~\cite{leo2021chiral} \hotspice{} allows each magnet to have a unique magnetic moment size $\mu_i$, temperature $T_i$ and energy barrier $E_{\mathrm{B},i}$. This enables, for instance, modeling some of the disorder due to lithographic variations by assigning a different shape anisotropy to each magnet, typically sampled from a Gaussian distribution with mean $E_\mathrm{B}$ and standard deviation $\sigma(E_\mathrm{B})$.~\cite{DisorderGroundStateASI} \par
\begin{figure*}[ht!]
    \includegraphics[width=\textwidth]{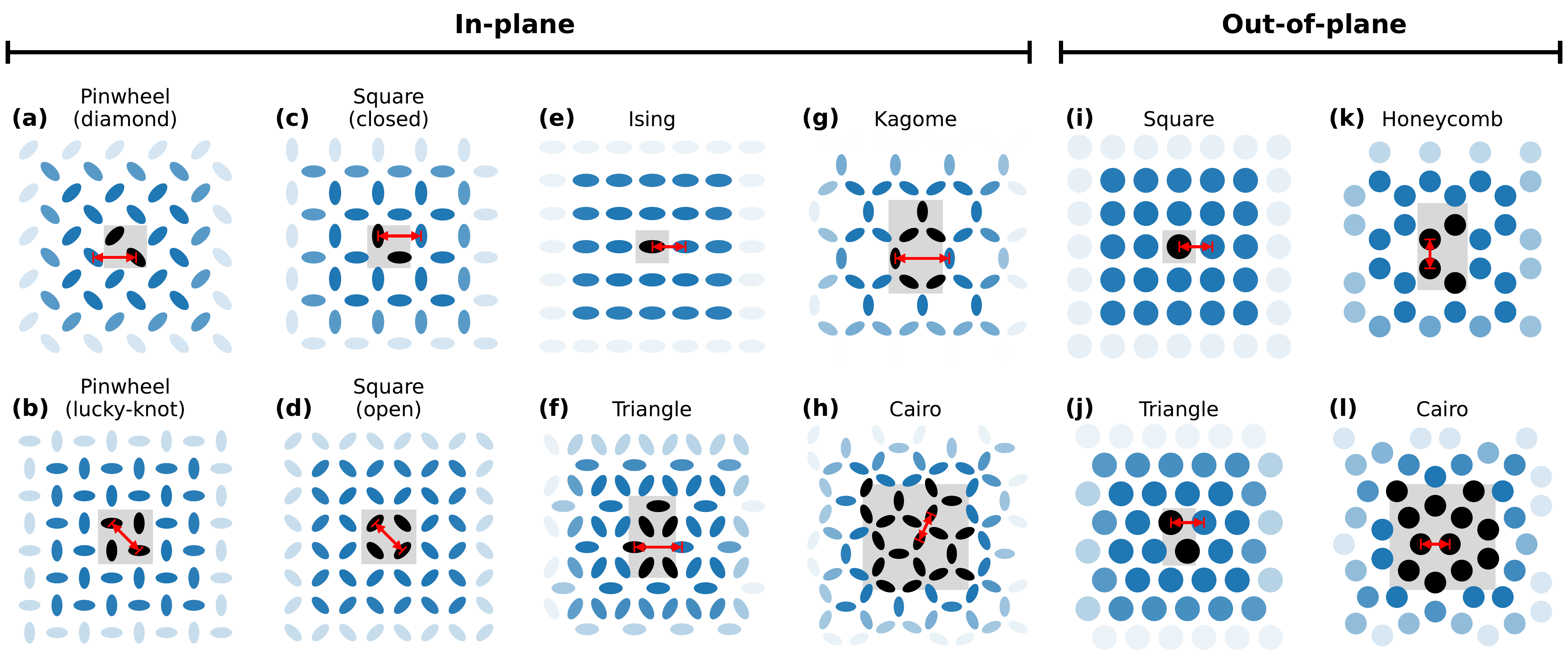}
    \caption{Predefined artificial spin ice (ASI) lattices available in \hotspice{}. The unit cell of each lattice is delineated by a central gray rectangle. The red indicator defines the lattice parameter $a$. In the Ising approximation, the magnetization of in-plane magnets (left) aligns along the major axis of the depicted ellipses. Out-of-plane magnets (right) are illustrated as circles.}
    \label{fig:ASIs}
\end{figure*}

Due to the periodic nature of many ASI lattices, \hotspice{} chooses to perform the simulation on a rectilinear grid, the benefits of which will be discussed in~\cref{sec:Implementation}. Each grid point may or may not contain a magnet, and the magnets must either all be IP or OOP. Many popular ASI lattices can be constructed in this manner, as showcased in~\cref{fig:ASIs}. Some lattices in this figure are related to each other: the magnets of the OOP lattices (i)-(l) are positioned at the vertices where magnets meet in the respective IP lattices (e)-(h), and the pinwheel lattices (a) and (b) are equal to the square lattices (c) and (d), respectively, but with each magnet rotated \ang{45}. \par 

\subsection{Energy calculation}
\label{sec:E}
\subsubsection{Energy terms}
\label{sec:E_terms}
\hotspice{} considers three energy terms\footnote{Additional energy terms can be defined by the user by inheriting from the \code{Energy} class.}: the magnetostatic interaction energy $E_{\mathrm{MS},i,j}$ between magnets $i$ and $j$, the Zeeman energy $E_{\mathrm{Z},i}$ of an external field $\vb{B}_\mathrm{ext}$ interacting with magnet $i$, and the exchange coupling energy $E_{\mathrm{exch},i,j}$ between nearest neighbors (NN) $i$ and $j$. By default, only the magnetostatic interaction is nonzero. For magnetic dipoles, they are calculated as follows:
\begin{align}
    E_{\mathrm{MS},i,j} &= \frac{\mu_0}{4 \pi} \ab(\frac{\vb{\mu}_i \bcdot \vb{\mu}_j}{\abs{\vb{r}_{ij}}^3} - \frac{3(\vb{\mu}_i \bcdot \vb{r}_{ij}) (\vb{\mu}_j \bcdot \vb{r}_{ij})}{\abs{\vb{r}_{ij}}^5}) \mathrm{,}
    \label{eq:E_MS} \\
    E_{\mathrm{Z},i} &= -\vb{\mu}_i \bcdot \vb{B}_\mathrm{ext} \mathrm{,} \label{eq:E_Z} \\
    E_{\mathrm{exch},i,j} &= J \frac{\vb{\mu}_i \bcdot \vb{\mu}_j}{\mu_i \mu_j} \mathrm{,} \label{eq:Eexch}
\end{align}
with $\vb{r}_{ij}$ the vector connecting the two magnetic dipoles $\vb{\mu}_i$ and $\vb{\mu}_j$, $\mu_0$ the vacuum permeability and $J$ the exchange coupling constant. 
The exchange interaction is usually not present in ASI ($J=0$), but can be relevant in cases such as interconnected ASI (whether by design or by lithographic inaccuracies); this was for example the case in~\ccite{Paper_PMA_ASI}, where one sample required accounting for the exchange interaction to achieve a proper fit. \par
The combined interaction energy $E_i$ of a single magnet $i$ is then given by
\begin{equation}
    E_i = E_{\mathrm{Z},i} + \sum_j E_{\mathrm{MS},i,j} + \sum_{j \in \mathcal{N}_i} E_{\mathrm{exch},i,j} \mathrm{,}
    \label{eq:Eint}
\end{equation}
with $\mathcal{N}_i$ the nearest neighbors of $i$.
Note that, due to the absence of magnetostatic self-energy in \cref{eq:Eint}, $E_i$ simply changes sign when magnet $i$ switches, and hence its switching energy $\Delta E_{i,1\rightarrow2} = -2 E_i$. 

\subsubsection{Finite-size corrections to the magnetostatic energy}
\label{sec:E_MS}
\begin{figure*}[ht!]
    \centering
    \includegraphics[width=\linewidth]{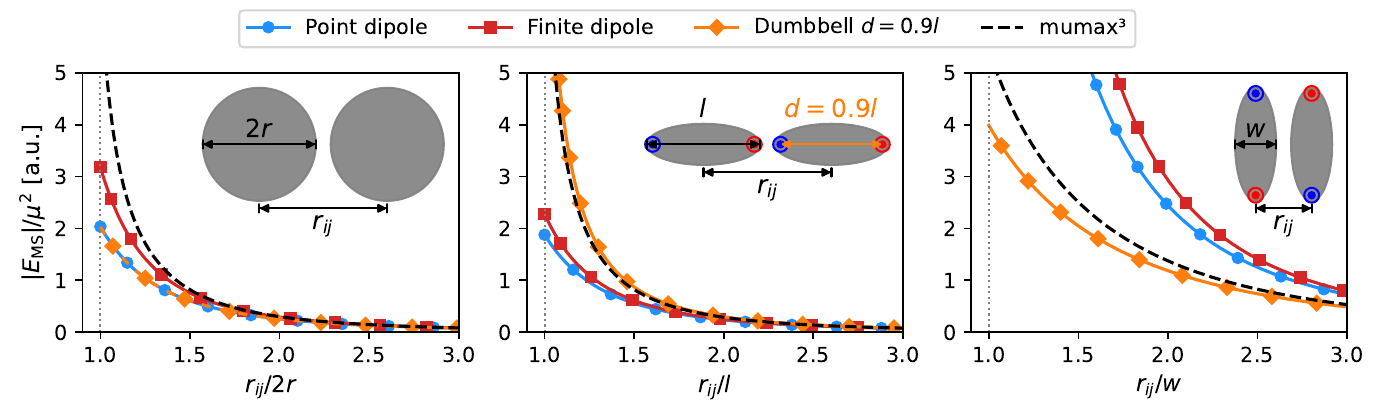}
    \caption{Magnitude of the magnetostatic interaction between two magnets as a function of their normalized center-to-center distance, for the three \hotspice{} calculation methods (point dipole, second-order correction for dipoles, and dumbbell) compared to a micromagnetic \mumax{} calculation. OOP magnets are assumed to be circular with diameter $2r$, IP magnets are ellipses with length $l$ and width $w=4l/11$. Positions of north and south magnetic charges used for the dumbbell calculation in IP magnets are shown as red {$\color{red}\odot$} and blue {$\color{blue}\odot$} dots and are a distance $d=0.9l$ apart within a magnet. The energy on the vertical axis was divided by $M^2$ to be independent of magnet volume.}
    \label{fig:DD_distance}
\end{figure*}
\cref{eq:E_MS,eq:E_Z,eq:Eexch} approximate each nanomagnet as a point dipole, but real nanomagnets have finite spatial extent. This does not affect the Zeeman and exchange energy since they are position-independent, and any shape-related effects can be captured by an appropriately chosen $\vb{B}_\mathrm{ext}$ and $J$. 
The magnetostatic interaction, however, depends on the relative position, orientation, and shape of all magnets. This may result in inadequate simulation of closely spaced ASI where the real magnetostatic coupling can be significantly stronger than predicted by a point dipole approximation. Therefore, two (mutually exclusive) improvements are proposed, which rescale the magnetostatic interaction energy between magnets.
\paragraph{Second-order correction for dipoles}
In~\ccite{Dipolar2Dparticles}, \textit{Politi and Pini} present a multipole expansion of the magnetostatic interaction, to account for the finite size of 2D nanomagnets (i.e., lateral dimensions $\gg$ thickness), assuming a uniform magnetization. This results in a second-order correction
\begin{equation}
    E_{\mathrm{MS},i,j} = E_{\mathrm{MS},i,j}^\mathrm{(0)} + E_{\mathrm{MS},i,j}^\mathrm{(2)} \mathrm{,}
\end{equation}
where $E_{\mathrm{MS},i,j}^\mathrm{(0)}$ is the original point dipole magnetostatic interaction given by~\cref{eq:E_MS}. The second-order correction is
\begin{align*}
    E_{\mathrm{MS},i,j}^\mathrm{(2)} = \frac{\mu_0}{4\pi} \frac{3\mathcal{I}_{ij}}{2} \Bigg[&3\frac{\vb{\mu}_i^\mathrm{OOP} \bcdot \vb{\mu}_j^\mathrm{OOP}}{r_{ij}^5} + \frac{\vb{\mu}_i^\mathrm{IP} \bcdot \vb{\mu}_j^\mathrm{IP}}{r_{ij}^5} \\ &-5\frac{(\vb{\mu}_i^\mathrm{IP} \bcdot \vb{r}_{ij}) (\vb{\mu}_j^\mathrm{IP} \bcdot \vb{r}_{ij})}{r_{ij}^7} \Bigg] \mathrm{,} \numberthis
\end{align*}
where $\vb{\mu}_i$ was split into its IP and OOP components. The shape of the nanomagnets is encapsulated in the single scalar $\mathcal{I}_{ij} = (\mathcal{I}_i + \mathcal{I}_j)/2$, which is calculated similar to a moment of inertia: $\mathcal{I}_i = \int_{\Omega_i} \abs{\vb{r} - \vb{r}_{0,i}}^2 d\vb{r}$ with $\vb{r}_{0,i} = \int_{\Omega_i} \vb{r} d\vb{r}$ the center of mass of magnet $i$. Assuming all magnets are elliptical cylinders with semi-major axis $a$ and semi-minor axis $b$, this reduces to $\mathcal{I}_{ij} = (a^2 + b^2)/4$. 
\paragraph{Dumbbell model}
Instead of representing a magnet as a point dipole, one may instead represent it as a pair of opposite magnetic charges.~\cite{MagneticMonopoles2008,MagneticMonopoleDynamics} This introduces a new parameter $d$: the effective distance between the north and south poles of a magnet, with respective positions $\vb{r_N}_i = \vb{r}_i + s_i\frac{d_i}{2}\vb{u}_i$ and $\vb{r_S}_i = \vb{r}_i - s_i\frac{d_i}{2}\vb{u}_i$. An appropriate choice of $d_i$ (slightly smaller than the physical length $l$ of the nanomagnet~\cite{DDG_Masterproef}) allows this dumbbell model to emulate the spatial extent of a real nanomagnet. The north and south poles are assigned the magnetic charges $+q_i$ and $-q_i$, respectively, with $q_i=\mu_i/d_i$.~\cite{MagneticMonopoles2008} \par 
The interaction energy between two magnetic charges $q$ and $q'$ can be derived from the magnetic version of Coulomb's law~\cite{ForceMagneticDipole} as
\begin{equation}
    E = -\int_\infty^{\vb{r}} \frac{\mu_0}{\num{4}\pi}\frac{qq'}{\abs{\vb{r}}^3} \vb{r} \cdot d\vb{r} = \frac{\mu_0}{\num{4}\pi} \frac{qq'}{\abs{\vb{r}}} \mathrm{.}
\end{equation}
The magnetostatic interaction energy between two nanomagnets is then the sum of their four mutual magnetic charge interactions, resulting in
\begin{align*}
    E_{\mathrm{MS},i,j} = \frac{\mu_0 \mu_i \mu_j}{4\pi d_i d_j} \Bigg(&\frac{1}{\abs{\mathbf{r_N}_i - \mathbf{r_N}_j}} + \frac{1}{\abs{\mathbf{r_S}_i - \mathbf{r_S}_j}}\\ - &\frac{1}{\abs{\mathbf{r_N}_i - \mathbf{r_S}_j}} - \frac{1}{\abs{\mathbf{r_S}_i - \mathbf{r_N}_j}}\Bigg) \mathrm{.} \numberthis \label{eq:E_MS_mono}
\end{align*}
The minus sign in the equation appears because the north and south poles have opposite charge.

\paragraph{Comparison}
The effect these two methods have on the magnetostatic interaction is illustrated in~\cref{fig:DD_distance}, as a function of the normalized distance between magnets. For out-of-plane (OOP) systems, the dumbbell model is inadequate due to the small fringe fields and the limited thickness of the magnets. Instead, the second-order dipole correction is more appropriate, yielding a significant improvement towards the ideal \mumax{} curve. \par
For IP systems, the dumbbell model constitutes a vast improvement over the standard point dipole treatment. The best correspondence with \mumax{} is found when the charge-to-charge distance $d$ is set slightly shorter than the length $l$ of a magnet, typically around $d/l\approx0.9$. This adjustment accounts for the curvature at the ends of most nanomagnets; similar values for $d/l$ were previously found in~\ccite{DDG_Masterproef} for typical nanomagnet shapes like ellipses and stadiums. In contrast, the second-order dipole correction has little effect in IP systems and can even increase the discrepancy from \mumax{}: it emulates increased spatial extent and therefore always increases the interaction, but in the anti-parallel configuration ($\uparrow \downarrow$) the point dipole model already overestimates the interaction. \par
Thus, the dumbbell model is preferred for IP systems, while the second-order dipole correction is most suitable for OOP systems.

\subsubsection{Effective energy barrier}
\label{sec:E_B}
To simulate ASI dynamics and hysteresis, it is crucial to know the height of the effective energy barrier $\widetilde{E_\mathrm{B}}$ which separates the two magnetization states of each magnet.\footnote{The effective barrier $\widetilde{E_\mathrm{B}}$ is distinct from the shape anisotropy $E_\mathrm{B}$: $\widetilde{E_\mathrm{B}}$ is a modification of $E_\mathrm{B}$ caused by the interaction with other magnets.} It can be calculated at varying levels of accuracy, which may result in different switching rates or even a different switching order.~\cite{leo2021chiral} \par
Recall that \cref{eq:Eint} is written in such a way that the interaction energy $E_i$ changes sign if the magnetization of either magnet is reversed: say $i$ reverses at time $t$, then $E_i(t^+) = -E_i(t^-)$. This formulation allows \hotspice{} to easily calculate the energy change $\Delta E_i = -2 E_i$ of magnet $i$ if it would switch (called its switching energy). \par
We will omit the subscript $i$ for the remainder of this section; it is implied that the following equations apply to each magnet individually.

\paragraph{Mean-barrier approximation}
The simplest way to approximate the effective energy barrier $\widetilde{E_\mathrm{B}}$ is to assume that the highest-energy state lies halfway between the two stable states. This means the barrier height changes at half the rate at which the switching energy $\Delta E$ changes, leading to the approximation $\widetilde{E_\mathrm{B}} = E_\mathrm{B} + \Delta E / 2$.~\cite{MC_TemperatureDesorption,DirectionalEnergyBarrier} However, this does not account for the extreme cases where the interactions are so strong that the energy barrier effectively disappears. This happens when $\abs{\Delta E}$ exceeds twice the shape anisotropy $E_\mathrm{B}$, leaving only one global minimum. To handle these situations, \hotspice{} calculates the effective energy barrier $\widetilde{E_\mathrm{B}}$ as follows:
\begin{equation}
    \widetilde{E_\mathrm{B}} = \begin{cases}
        E_\mathrm{B} + \frac{\Delta E}{2} & \text{if } \abs{\frac{\Delta E}{2}} < E_\mathrm{B}, \\
        \Delta E & \text{otherwise}.
    \end{cases}
\end{equation}
This way, $\Delta E$ serves as the barrier when the original energy barrier disappears.

\paragraph{Asymmetric barrier (IP only)}
The simple approximation above is insufficient for many in-plane ASI. In real nanomagnets, the magnetization must rotate for switching to occur: either clockwise ($\circlearrowright$) or anticlockwise ($\circlearrowleft$).~\cite{DirectionalEnergyBarrier}
In an asymmetrical environment (when the effective field has a nonzero component perpendicular to the easy axis) one of these two rotation directions will be preferred.~\cite{leo2021chiral,DirectionalEnergyBarrier} Take for example the pinwheel ASI: any two neighboring magnets form a T-shape, so the magnet pointing into the side of the other will greatly influence whether the other prefers $\circlearrowleft$ or $\circlearrowright$ rotation.~\cite{DirectionalEnergyBarrier} Accounting for the existence of these separate chiral switching channels profoundly affects the switching rates and transition kinetics, since switching will occur predominantly via the more favorable pathway.~\cite{leo2021chiral} \par
In the dipole model, this can be accounted for by keeping track of the energy of each magnet in these transitional `perpendicular' states. We therefore introduce $E_\perp$, which represents the energy of a magnet if it would point \ang{90} counterclockwise from its normal magnetization direction, but without taking into account the shape anisotropy ($E_\mathrm{B}$) associated with that orientation.
(Note: the bistability of the magnets does not change; we are simply putting `test dipoles' in the perpendicular orientation, but the magnets are never actually put in these states.) 
The equation for the effective barrier along the two rotation pathways ($\pm$) then becomes
\begin{equation}
    \widetilde{E_\mathrm{B}} = \begin{cases}
        E_\mathrm{B} \pm \rho E_\perp + \frac{\Delta E}{2} & \text{if } \abs{\frac{\Delta E}{2}} < E_\mathrm{B} \pm \rho E_\perp, \\
        \Delta E & \text{otherwise}, \\
    \end{cases}
\end{equation}
which results in two different barriers if $E_\perp \neq 0$. Note that the value of each term in this equation can differ for each magnet in the ASI. 
The parameter $\rho = \mu_\perp/\mu_\parallel > 0$ was introduced to account for non-coherent magnetization reversal processes like domain wall nucleation and propagation, which result in an effective reduction of the magnetic moment during reversal.~\cite{leo2021chiral,TimeResolvedDynamicsSOT} Using a value $\rho < 1$ improves correspondence with experimental observations, as we will show in~\cref{sec:pinwheel_reversal}. \par

\paragraph{Exact solution (OOP only)}
If an analytical expression for the energy as a function of magnetization angle $\theta$ relative to the magnet's easy axis is known, then $\widetilde{E_\mathrm{B}}$ can be calculated exactly. The shape anisotropy creates a basic energy profile with two minima at $\theta=0$ and $\theta=\pi$, but the exact form of this profile depends on the magnet's shape; for ellipsoidal magnets, it is $\frac{-E_\mathrm{B}}{2} \cos{2\theta}$.~\cite{neel1949theorie} Assuming a uniform magnetization in each magnet\footnote{Only perfectly ellipsoidal magnets have uniform magnetization in a uniform external field.~\cite{EllipsoidDemag,MaxwellElectricityMagnetism}}, the magnetostatic and Zeeman interactions add a term proportional to $\cos(\theta - \phi)$, with $\phi$ the angle of their combined effective field. Thus, the total profile is a sum of two sines and can be fully characterized if $E$ and $E_\perp$ are known. However, in the general case, this results in a transcendental equation requiring numerical approximation to solve, which is not done in \hotspice{} for performance reasons. \par
In OOP ASI, an explicit expression can still be obtained because the effective field aligns with the easy axis, making $E_\perp=0$. This leads to a quadratic relation as described in~\ccite{StonerWohlfarth2008}:
\begin{equation}
    \widetilde{E_\mathrm{B}} = \begin{cases}
        E_\mathrm{B} \ab(\frac{\Delta E}{4 E_\mathrm{B}} + 1)^2 & \text{if } \abs{\frac{\Delta E}{2}} < E_\mathrm{B}, \\
        \Delta E & \text{otherwise}. \\
    \end{cases}
\end{equation}

\subsection{Dynamics}
\label{sec:dynamics}
A magnet may spontaneously switch to the opposite magnetization state due to thermal fluctuations. The time evolution of the ASI is evaluated in a stepwise manner by an update algorithm that determines which magnet should switch next. The two distinct types of kinetic Monte Carlo (KMC) algorithms---rejection-free KMC and rejection KMC---were both implemented in \hotspice{}. In the context of ASI, we refer to these as \textit{N\'eel-Arrhenius switching}
and \textit{Metropolis-Hastings}, respectively.~\cite{PhysicalTimeKMC} The former is more suitable for simulating the temporal evolution of the system, while the latter can be used to sample the equilibrium distribution of the state space. For a broader overview of Monte Carlo methods, we refer to Ref.~\cite{IntroductionMC}, which may also clarify the at times confusing naming present throughout literature.

\subsubsection{N\'eel relaxation: temporal evolution}
N\'eel relaxation theory~\cite{neel1949theorie} states that, for an isolated nanomagnet, the switching rate $\nu$ is given by the N\'eel-Arrhenius equation
\begin{equation}
    \nu = \nu_0 \exp\ab(-\frac{E_\mathrm{B}}{\kBT}) \mathrm{,}
    \label{eq:Néel}
\end{equation}
with $\kBT$ the thermal energy and $\nu_0$ the attempt frequency. An estimate of $\nu_0$ for coherent magnetization reversal can be obtained from the limit $E_\mathrm{B} \rightarrow 0$. The gyromagnetic precession frequency of the magnetization of a nanomagnet is on the order of \qtyrange{e9}{e10}{\hertz}.~\cite{BrownThermalFluctuations,bean1959superparamagnetism}
As $E_\mathrm{B} \rightarrow 0$, the switching rate $\nu$ should approach this value, and in this limit \cref{eq:Néel} implies that $\nu \rightarrow \nu_0$, so we use $\nu_0=\qty{e10}{\hertz}$.~\cite{JM_Masterproef} 
An order-of-magnitude estimate of $\nu_0$ suffices, because any small (i.e., $\sim \kBT$) change of $E_\mathrm{B}$ will translate to an exponential change in switching rate. \par 
For mutually interacting magnets, an adjusted version of \cref{eq:Néel} can be used where $E_\mathrm{B}$ is replaced by the effective energy barrier $\widetilde{E_\mathrm{B}}$. In the general case where the energy barriers for clockwise and anticlockwise rotation during switching differ, these two switching channels $(\circlearrowright, \circlearrowleft)$ will separately follow~\cref{eq:Néel}, so their switching frequencies must be combined. This yields the total switching rate presented in~\cite{DirectionalEnergyBarrier}:
\begin{equation}
    \nu = \nu_\circlearrowleft + \nu_\circlearrowright = \frac{\nu_0}{2} \ab[\exp\ab(-\frac{\widetilde{E_\mathrm{B}}{}_{,\circlearrowleft}}{\kBT}) + \exp\ab(-\frac{\widetilde{E_\mathrm{B}}{}_{,\circlearrowright}}{\kBT})] \mathrm{,}
    \label{eq:Néel_2}
\end{equation}
where we assigned a halved attempt frequency $\nu_0/2$ to either switching channel such that~\cref{eq:Néel_2} reduces to~\cref{eq:Néel} in the case of $\widetilde{E_\mathrm{B}}{}_{,\circlearrowleft}=\widetilde{E_\mathrm{B}}{}_{,\circlearrowright}$.~\cite{leo2021chiral} \par
One iteration of the algorithm is as follows:
\begin{enumerate}
    \item Calculate the switching rate $\nu_i$ of all magnets (i.e., $\forall i$) based on the interaction energies $E_i$, and hence effective energy barriers $\widetilde{E_{\mathrm{B},i}}$, present in the current magnetization state.
    \item Generate a random switching time interval $\Delta t_i$ for each magnet $i$, sampled from an exponential distribution with mean value $1/\nu_i$.
    \item Determine which magnet $j$ has the smallest such time $\Delta t_j = \min_i \Delta t_i$.
    \item To prevent excessively long switching times, a parameter $t_\mathrm{max}$ was introduced.
    \begin{itemize}
        \item \textit{If $t + \Delta t_j \leq t_\mathrm{max}$}: increment the elapsed time $t$ by $\Delta t_j$ and switch magnet $j$.
        \item \textit{If $t + \Delta t_j > t_\mathrm{max}$}: increment the elapsed time $t$ by $t_\mathrm{max}$ without switching a magnet.
    \end{itemize}
\end{enumerate} 
The maximum time $t_\mathrm{max}$ (default value of one second) prevents the simulation from advancing too far into the future, as the exponential character of the N\'eel-Arrhenius law can cause switching times to become much longer than what could ever be observed experimentally. It can also be used to apply time-dependent external fields: for example, when a sinusoidal signal of frequency f is applied to the lattice $t_{\mathrm{max}}=20/f$ ensures the waveform is captured in sufficient detail. \par

\subsubsection{Metropolis-Hastings: sampling equilibrium states}
Metropolis-Hastings is a rejection-based kinetic Monte Carlo (KMC) method designed to sample the state space at thermal equilibrium, where the probability of each state appearing is proportional to their Boltzmann factor.~\cite{IntroductionMC,kyimba2006comparisonIsingAlgorithms} Hence, in contrast to N\'eel-Arrhenius switching, Metropolis-Hastings is not intended to accurately model the system's transient dynamics and is instead more suitable for examining equilibrium statistical quantities in ASI, like the average magnetization, heat capacity, correlation...~\cite{ApparentFMpinwheel} \par
The algorithm repeats the following steps:
\begin{enumerate}
    \item Select a magnet $i$ at random (with all magnets equally likely to be chosen).
    \item Calculate the energy change $\Delta E_i$ if this magnet were to switch.
    \item Switch the magnet with probability
        \begin{equation}
            P_i = \begin{cases}
                \exp(-\Delta E_i/\kBT), & \text{if } \Delta E_i > 0 \mathrm{,} \\
                1 & \text{otherwise} \mathrm{.}
            \end{cases}
        \end{equation} 
    \item \textit{Optional}:
        Increment the elapsed time $t$ by
        \begin{equation}
            \Delta t = -\frac{\exp\ab(\widetilde{E_\mathrm{B}}\big/\kBT\ab) \ln{\chi}}{N \nu} \mathrm{,}
            \label{eq:Metropolis_time}
        \end{equation}
        with $N$ the number of magnets in the system and $\chi$ a uniformly distributed random variable in $(0,1]$.~\cite{PhysicalTimeKMC}
\end{enumerate}
This algorithm satisfies detailed balance and ergodicity, thus ensuring equilibrium is eventually reached. However, the rate of convergence may vary as noted in~\cref{sec:OOP_Exchange}~\cite{jang2004stochastic}: end users must take care to perform sufficient Monte Carlo steps to ensure thermalization. For enhanced performance, multiple sufficiently distant magnets can be selected simultaneously, see~\cref{sec:multiswitch}. \par
Whereas N\'eel relaxation relies on an explicit calculation of the elapsed time, Metropolis-Hastings does not strictly require this knowledge. Hence, the last step of the algorithm where the elapsed time is calculated, is optional. For a long time, the notion of a well-defined elapsed time in rejection KMC was controversial.~\cite{nfoldMCalgorithm,GlauberTimescale_sadiq1984,MCSim_StatPhys} 
Often, the number of MC steps per site was used as a crude measure of `elapsed time', but eventually a formal derivation for the physical time scale in rejection KMC was presented in~\ccite{PhysicalTimeKMC} (\cref{eq:Metropolis_time}). Note that the effective energy barrier $\widetilde{E_\mathrm{B}}$ only appears in the elapsed time: it does not influence the switching probability in the Metropolis-Hastings algorithm because it has no effect on the equilibrium state.

\subsection{Implementation details}
\label{sec:Implementation}
\hotspice{} represents an ASI as a rectilinear grid of non-uniform unit cells, with magnets positioned at selected grid points. In simulating an ASI, we faced a trade-off between the freedom to place magnets arbitrarily and the efficiency of calculation. We opted to prioritize efficiency and accept the geometrical restriction, as most ASI research focuses on periodic lattices. Despite the seemingly restrictive nature of the rectilinear grid,~\cref{fig:ASIs} illustrates its versatility in forming various periodic lattices, with only the Cairo lattices requiring grid non-uniformity. Real-time visualization is simple and efficient for a rectilinear grid, as the underlying matrix can directly be cast to a pixel image. \par
By leveraging the unit cell concept in periodic lattices and the efficient indexation of a rectilinear grid in computer memory, several aspects of the calculation can be performed more efficiently than for free-form ASI. The unit cell of each lattice in~\cref{fig:ASIs} is depicted as a gray rectangle. Although non-rectilinear unit cells with fewer magnets could be identified for some lattices, \hotspice{} does not consider these to reduce complexity and maintain a clear connection to the underlying rectilinear grid of the ASI implementation.


\subsubsection{Kernels}
Pre-calculated ``kernels'' are used to efficiently update the magnetostatic interaction after each switch. For each magnet $i$, a kernel $\vb{K}_i(j)$ stores the strength of this interaction between itself and all other magnets $j$, enabling the quick calculation of their magnetostatic interaction energy as $E_{\mathrm{MS},i,j}=\vb{K}_i(j) s_i s_j$ by precalculating all $\vb{K}_i(j)$ values once when the ASI is created. However, for large arrays, this approach becomes impractical due to the need to store $\mathcal{O}(N^2)$ elements $\vb{K}_i(j)$ if the underlying rectangular grid has size $N = L_x \times L_y$. \par
By leveraging unit cells, this storage requirement can be reduced to $\mathcal{O}(N)$. Labeling each spot in the unit cell by an index $q$, each magnet in the lattice corresponds to an index $q$. The surrounding magnets of a magnet at spot $q$ will always have the same layout, except for a different cutoff at the border if PBC are disabled. Thus, all possible interactions a magnet at site $q$ experiences can be stored in a single $(2L_x-1) \times (2L_y-1)$ matrix $\vb{K}_q$, with the magnet $q$ at the center of the matrix and the element $\vb{K}_q(L_x + x, L_y + y)$ representing the interaction between a magnet with index $q$ at site $(v,w)$ and the magnet at site $(v+x, w+y)$. This method requires storing only a few kernels---one for each magnet in the unit cell---rather than one for each magnet in the entire lattice, thereby reducing the number of stored kernels by $\mathcal{O}(N)$. \par
Note that, if the asymmetric energy barrier is to be accounted for, two additional kernels are needed for each magnet in the unit cell: one where the central magnet is rotated \ang{90} (for initial calculation of $E_\perp$), and one where all other magnets are rotated \ang{90} (for updating $E_\perp$ when a magnet switches). \par
The grid enables the straightforward implementation of first-order PBC by adding eight offset versions of a kernel to itself, which does not impact performance. A cut-off radius for the magnetostatic interaction can also be imposed by setting the relevant kernel values to zero.

\subsubsection{Multi-switching in Metropolis-Hastings}
\label{sec:multiswitch}
Since the Metropolis-Hastings algorithm is designed primarily for sampling equilibrium states rather than capturing the temporal evolution of the system, a straightforward performance improvement can be achieved by selecting multiple magnets simultaneously rather than sequentially. This allows for better usage of the parallel processing capabilities of the GPU, as the magnetostatic energy can then be updated using a convolution. However, to avoid issues with simultaneous switching of nearby magnets, we enforce a minimum distance $r$ between selected magnets. The criterion we use is that two simultaneously sampled magnets should never affect each other's switching probability by more than a user-adjustable factor $Q \in~]0,1[$ (commonly set to 0.01). This leads to the following expression for $r$:
\begin{equation}
    r \geq \sqrt[3]{\frac{2 \mu_0 \max_i \mu_i^2}{\pi Q \kBT}} \mathrm{,}
\end{equation}
valid for the Metropolis-Hastings switching probability $P(\Delta E) = \min(1, \exp(-\Delta E/\kBT))$ and with $\max_i \mu_i^2$ the square of the largest magnetic moment in the lattice. \par
For magnet selection, we employed a modified ``Stratified Jittered Grid'' 
algorithm. The simulation domain is divided into subregions of $R_x \times R_y$ grid cells, with $R_x \geq 2$ and $R_y \geq 2$ chosen such that each subregion has a physical size of at least $r \times r$. A quarter of these subregions are chosen such that they are non-adjacent, and a single magnet is sampled from each of them.~\cite{SamplingPolyominoes} This ensures that sampled magnets are at least a distance $r$ apart, though their average distance will be $2r$. 
Although this method is less random, dense and versatile than Poisson Disk Sampling~\cite{PoissonDiskComparison,FastPoissonDiskSampling,SamplingPolyominoes}, it is efficient to compute in parallel\footnote{\ccite{PoissonDiskParallel} presents a parallel Poisson Disk algorithm which generates samples over only a few iterations, though it is nontrivial to restrict the samples to a non-uniform grid without violating the minimal distance constraint.} and synergizes well with our grid-based ASI implementation. \par
The ``supergrid'' of subregions is slightly smaller than the ASI to prevent PBC from violating the minimal distance requirement and is randomly shifted over the simulation domain to ensure uniform coverage and proper behavior near the edges. Note that PBC can make it impossible to maintain a spacing $>r$ in very small systems, or with small $Q$ values; in such cases, we resort to selecting a single magnet. \par 

\subsubsection{Simulation output}
The results of a simulation are accessible through the attributes of the ASI object used for computation. Following the philosophy of Python package design, the processing and storage of these attributes during and after simulations is mostly left to the end user. Due to the underlying grid, most of these attributes are stored as 2D arrays. Most importantly, the magnetization state $s_i = \pm 1$ of all magnets can be accessed in this manner. The simulation time, measured by the elapsed time or MC steps, is tracked automatically. The grid also enables straightforward calculation of e.g. spatial correlations by convolving these arrays with appropriate masks.

\section{Verification} 
\label{sec:Verification}
In this section, the correct implementation of basic aspects of the model used by \hotspice{} is verified, by comparing simulations to analytical solutions for several systems of varying complexity. The Metropolis-Hastings algorithm is used, thereby verifying its ability to sample the equilibrium state space.
\subsection{Exchange-coupled OOP square system}
\label{sec:OOP_Exchange}
\begin{figure}
    \centering
    \includegraphics[width=\columnwidth]{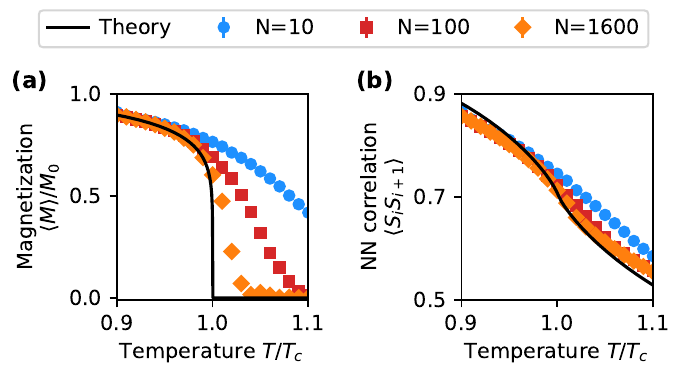}
    \includegraphics[width=\columnwidth]{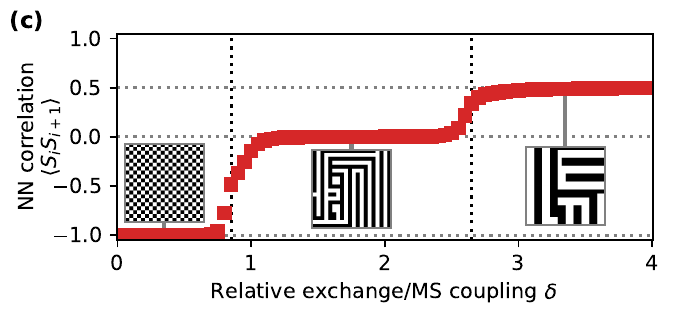}
    \caption{\hotspice{} simulation of a square-lattice exchange-coupled Ising system using the Metropolis-Hastings algorithm. (a) Average magnetization and (b) nearest-neighbor (NN) correlation as a function of temperature. Discrepancies due to critical slowing down above $T_c$ improve with more Monte Carlo steps per site $N$. (c) NN correlation when including long-range magnetostatic interactions, as a function of relative NN magnetostatic/exchange coupling $\delta$. Transitions occur at $\delta=0.85$ and 2.65, indicated by dotted lines. Insets show the magnetization state with growing stripe domains: white corresponds to spin `up', black to `down'.}
    \label{fig:Verification_OOP}
\end{figure}
The 2D square-lattice exchange-coupled Ising model is one of few exactly solvable systems in statistical physics.~\cite{ExactlySolvedModelsStatMech} Its average magnetization $M$ is temperature-dependent:
\begin{equation}
    M = \sqrt[8]{1-\sinh^{-4}(2J/\kBT)} \mathrm{,}
    \label{eq:OOP_exchange_M}
\end{equation}
with $J$ the exchange coupling constant.~\cite{Correlations2DIsing,IsingSpontaneousMagnetization,coey2010magnetism}
This system exhibits a second-order phase transition at the critical temperature $T_c=\frac{2J}{k_\mathrm{B}\ln(1 + \sqrt{2})}$.~\cite{ExactlySolvedModelsStatMech}
The nearest-neighbor correlation can also be calculated analytically:
\begin{equation}
    \langle S_i S_{i+1} \rangle = 
    \begin{cases}
        \sqrt{1+k} \ab[\frac{1-k}{\pi}K(k) + \frac{1}{2} \ab] &\text{for } T < T_c \mathrm{,} \\ 
        \sqrt{1+k} \ab[\frac{1-k}{\pi k}K(1/k) + \frac{1}{2} \ab] &\text{for } T > T_c \mathrm{.} \\ 
    \end{cases}
\end{equation}
with $K$ the complete elliptic integral of the first kind and $k=1/\sinh^2(2J/\kBT)$.~\cite{Correlations2DIsing}

The result from the \hotspice{} simulation of this Ising system is shown in \crefSubFigRef{fig:Verification_OOP}{a-b}, as calculated for an $800 \times 800$ lattice on GPU with maximal Metropolis-Hastings multi-sampling ($Q=+\infty$). The system was not reset between successive temperature steps because the theoretical curves are monotonously decreasing.~\cite{MCinStatPhys} The \hotspice{} result corresponds well to theory below $T_c$ and in the high-temperature limit. Just above $T_c$, however, the average magnetization only slowly evolves to the expected value. This is a symptom of the well-known phenomenon called ``critical slowing down'', which originates from a divergence in the autocorrelation time $\tau$ near the critical point, causing subsequent Monte Carlo configurations to be highly correlated.~\cite{NumericalDynamicalNiedermayer,CompStatPhys,StatisticalMechanicsAlgorithmsComputations} As a result, the system explores the phase space very slowly, particularly with single-spin flip algorithms like Metropolis-Hastings.~\cite{StatisticalMechanicsAlgorithmsComputations}
Although cluster algorithms like the Wolff algorithm~\cite{Wolff} can mitigate this effect, they are not intended for application beyond the 2D Ising system.

\subsection{Exchange- and magnetostatically coupled OOP square system}
Including long-range magnetostatic interactions into an exchange-coupled square-lattice Ising system significantly alters its behavior, which is then determined by the ratio $\delta = E_{\mathrm{exch},i,j}/E_{\mathrm{MS},i,j}$ ($j \in \mathcal{N}_i$) representing the balance between the exchange coupling and magnetostatic interaction. Analytical predictions remain possible: for $\delta < 0.85$, the magnetostatic coupling dominates, leading to a checkerboard state. As $\delta$ increases, the ever-stronger exchange coupling leads to the formation of ferromagnetic domains, which organize into stripes due to the magnetostatic interaction, with the stripe width determined by $\delta$.~\cite{StripedDipolarIsing} \par
The average stripe width is reflected in the NN correlation $\langle S_i S_{i+1} \rangle$, as shown in~\crefSubFigRef{fig:Verification_OOP}{c} for a \hotspice{} simulation. Consistent with the theoretical predictions of~\ccite{StripedDipolarIsing}, for $\delta < 0.85$ a checkerboard state exists with $\langle S_i S_{i+1} \rangle = -1$. In the range $0.85 < \delta < 2.65$, stripe domains with a width of 1 row are preferred, leading to $\langle S_i S_{i+1} \rangle = 0$. Beyond $\delta=2.65$, a 2-row width becomes preferable with $\langle S_i S_{i+1} \rangle = 0.5$. Increasing $\delta$ further leads to ever wider stripe domains, and in the limit $\delta \rightarrow +\infty$, the correlation approaches $\langle S_i S_{i+1} \rangle \rightarrow 1$. \par 

\subsection{Non-interacting spin ensemble}
When an external field of magnitude $B$ is applied to a non-interacting ensemble of Ising spins, the average magnetization follows the relation $\langle M \rangle / M_0 = \tanh(\mu B/\kBT)$, as follows directly from the partition function. \crefSubFigRef{fig:Verification_IP}{a} demonstrates that \hotspice{} correctly reproduces the expected result.

\subsection{Square-to-pinwheel transition angle}
\begin{figure}
    \centering
    \includegraphics[width=0.42\columnwidth]{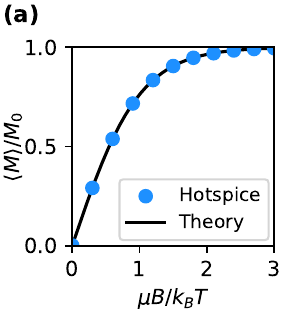}
    \includegraphics[width=0.56\columnwidth]{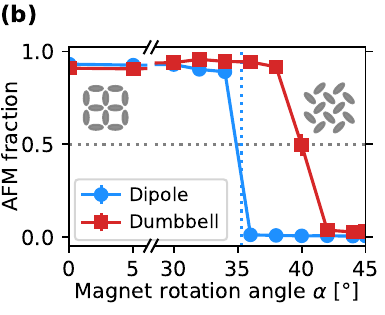}
    \caption{\hotspice{} simulations using the Metropolis-Hastings algorithm. (a) Average magnetization of a non-interacting ensemble of spins as a function of the applied field $B$. (b) Fraction of vertices with net zero magnetization at equilibrium, as square ASI (left) transitions to pinwheel ASI (right) by rotating individual magnets. The theoretical transition angle $\alpha_c \approx \ang{35.3}$ for the dipole model is indicated by the vertical dotted line. The dumbbell model uses a charge-to-charge distance $d=\SI{220}{\nano\metre}$.}
    \label{fig:Verification_IP}
\end{figure}
\label{sec:Verification_IP_SquarePinwheel}
The in-plane square and pinwheel ASI lattices can be continuously transformed into each other by rotating each individual magnet by \ang{45}. Their ground state magnetic ordering differs significantly: square ASI has an antiferromagnetic (AFM) ground state, where all vertices have a net zero magnetization, while pinwheel ASI exhibits superferromagnetic order, where all magnets of the same orientation are magnetized in the same direction.~\cite{ApparentFMpinwheel} Therefore, a critical angle $\ang{0} < \alpha_c < \ang{45}$ must exist where the ground state transitions between these two extremes. For the dipole model, theoretical calculations predict this transition at $\alpha_c = \arcsin(\sqrt{3}/3) = \ang{35.3}$.~\cite{AFM-FM-transition-Pinwheel,MagicAngle} For the dumbbell model, the transition angle depends on the distance $d$ between the magnetic charges within a magnet, but is always larger than for the dipole model.~\cite{AFM-FM-transition-Pinwheel}  \par 
The results of a \hotspice{} simulation are shown in \crefSubFigRef{fig:Verification_IP}{b}. To quantify this transition, we measure the fraction of vertices with net zero magnetization -- this value is 0 for superferromagnetic order while it is 1 for AFM order. We used the same lattice compression as described in~\ccite{AFM-FM-transition-Pinwheel}, making our lattice spacing $a=\SI{240}{\nano\metre} / \sin(\ang{45}+\alpha)$ angle-dependent: it varies from \SI{340}{\nano\metre} at $\alpha=\ang{0}$ to \SI{240}{\nano\metre} at $\alpha=\ang{45}$. For the dipole model, the transition occurs at $\approx \ang{35}$ as expected, while the dumbbell model (here with $d = \SI{220}{\nano\metre}$) indeed transitions at a larger rotation angle. 

\section{Applications}

To explore the accuracy of the various model variants, we compare simulations to several specific experiments. The pinwheel hysteresis highlights the importance of the asymmetric energy barrier, while the kagome reversal is an example where the dumbbell model captures a key detail of the reversal process which could not be reproduced by dipole models. We finish with a demonstration of ``clocking'' in pinwheel ASI, which has applications in reservoir computing.~\cite{clocking-protocol} Python codes for these simulations are provided in the supplementary information.

\begin{figure*} 
    \centering
    \includegraphics[width=\textwidth]{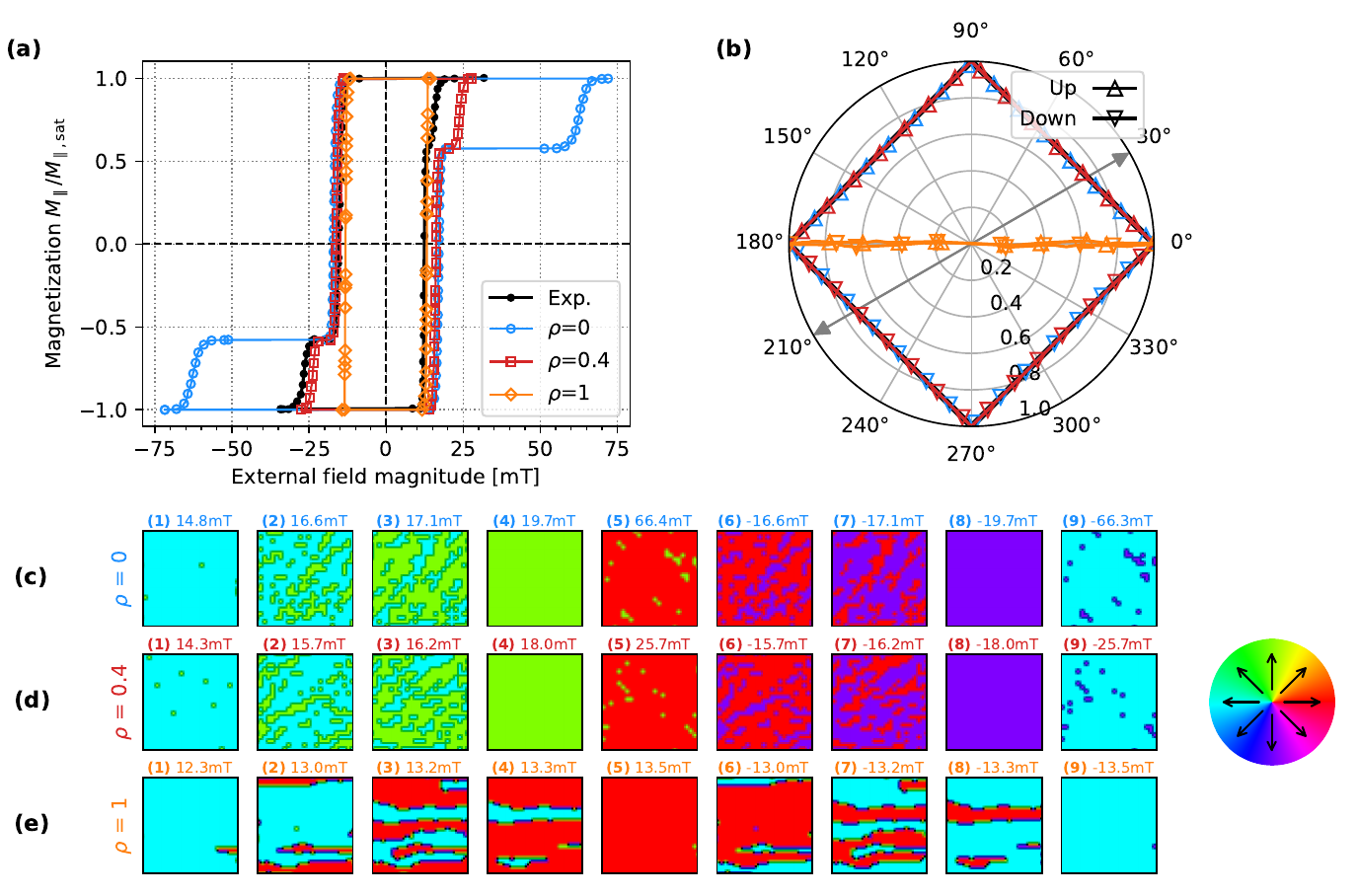}
    \caption{Hysteresis of the pinwheel ASI for an applied field at \ang{30} with respect to the system edges. The parameter $\rho = \mu_\perp/\mu_\parallel$ modulates the effect of the asymmetric barrier described in~\cref{sec:E_B}.
    \textbf{(a)} Component of the average magnetization along the direction of the external magnetic field. The experimental hysteresis by Li \textit{et al.}~\cite{li2018pinwheel} is shown in black. 
    \textbf{(b)} In-plane path of the average magnetization vector. Triangles pointing up (down) correspond to the ramp-up (down) of the external field. The external field direction is indicated by the double-headed arrow.
    \textbf{(c-e)} Snapshots of the system when passing through threshold values of $M_\parallel/M_{\parallel,\mathrm{sat}}$, in chronological order through the hysteresis loop. The color of each pixel shows the average magnetization angle of the magnets in a small area of the system, encoded according to the color wheel shown on the right. Snapshots (1,5,9) are taken near the saturated $M_\parallel/M_{\parallel,\mathrm{sat}} \approx \pm 1$ state, (3) and (7) at the zero-average $M_\parallel/M_{\parallel,\mathrm{sat}} = 0$, and (2,4,6,8) at $M_\parallel/M_{\parallel,\mathrm{sat}} = \pm\sqrt{3}/3 \approx 0.58$, which is the magnetization at the plateau between the two steps for $\rho=0$ and $\rho=0.4$.}
    \label{fig:test_pinwheelReversal}
\end{figure*}
\subsection{Pinwheel reversal} 
\label{sec:pinwheel_reversal}

Pinwheel ASI (\crefSubFigRef{fig:ASIs}{a-b}) can be seen as consisting of two intertwined sublattices whose magnets are perpendicular to each other. The ground state of this system consists of superferromagnetic domains, where all magnets within each sublattice are magnetized in the same direction.~\cite{EmergentChiralityRatchet,ApparentFMpinwheel,RC_ASI} \par
Li \textit{et al.}~\cite{li2018pinwheel} performed an experimental study to observe the reversal of `diamond'-edge pinwheel ASI (\crefSubFigRef{fig:ASIs}{a}) under an applied external field. The system exhibits hysteresis: a strong field drives it towards a uniform state, where it remains when the field is removed since this is the ground state of pinwheel ASI. Notably, when Li~\textit{et al.} applied the field at an oblique angle (\ang{30}) to the ASI edges, the reversal occurred in two distinct steps: the sublattice which was more aligned with the field (\ang{15} to the easy axis) reversed first, followed by the second sublattice for which the field was at \ang{75}. Accurate simulation of the second reversal step will require accounting for the asymmetry in the energy barrier between clockwise and counterclockwise switching, due to this near-perpendicular field for the second sublattice. \par
We replicated this experiment using \hotspice{} for a few values of $\rho=\mu_\perp/\mu_\parallel$ to clearly observe the effect of the asymmetric barrier, resulting in the hysteresis loops shown in~\cref{fig:test_pinwheelReversal}. The N\'eel update algorithm was used, because the Metropolis-Hastings algorithm samples the equilibrium state space and therefore would not capture the hysteresis observed experimentally -- a hysteresis only exists because the equilibrium is not reached within the timescale of the observation. 
We used parameters derived from the experimental configuration when possible: an ASI of $25 \times 25$ unit cells with a NN center-to-center distance of \qty{420}{\nano\meter} was used, at room temperature (\qty{300}{\kelvin}). The magnetic moment and energy barrier of each magnet can be derived from the geometry of the $470\times170\times\qty{10}{\nano\meter}$ stadium-shaped magnets and the permalloy (\ce{Ni_80 Fe_20}) saturation magnetization $M_\mathrm{sat}=\qty{800}{\kilo\ampere\per\meter}$. This yields a magnetic moment $\mu = M_\mathrm{sat} V = \qty{5.9e-16}{\ampere\meter\squared}$, and a \mumax{}~\cite{MuMax3} simulation reveals that the energy barrier of such magnets is $\approx \qty{60}{\electronvolt}$. These values result in a good match with the experimental field magnitude for the first reversal step. \par
\begin{figure*}[!ht]
    \centering
    \includegraphics[width=\textwidth]{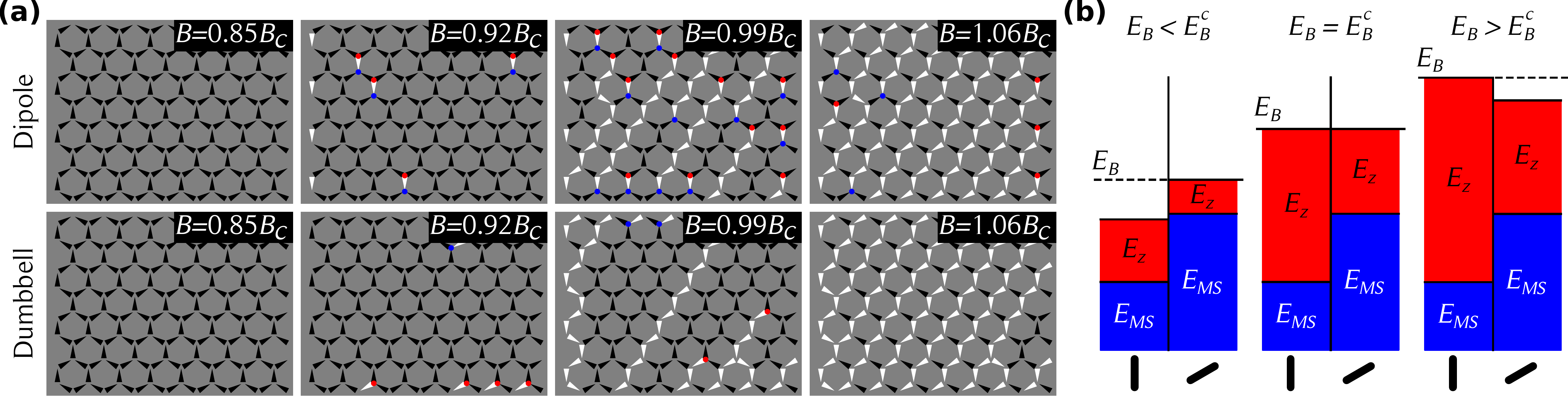}
    \caption{Reversal of a kagome ASI under an external field applied at \ang{-93.6} relative to the horizontal axis. (a) Comparison between the dumbbell and dipole models at different external field magnitudes $B$. Note that the critical field $B_C$ differs between the models:  $B_C = \qty{26.7}{\milli\tesla}$ for the dumbbell model and $B_C = \qty{23.9}{\milli\tesla}$ for the dipole model. The NN center-to-center distance is \qty{500}{\nano\meter}, with a length $d=\qty{470}{\nano\meter}$ used in the dumbbell model. The magnets have a magnetic moment $\mu=\qty{1.1}{\femto\ampere\meter^2}$, and their energy barrier is normally distributed with a mean of \qty{120}{\electronvolt} and \qty{5}{\percent} standard deviation. Arrows represent the magnetic moment of the magnets, and the blue and red points ({\color{blue}\textbullet} and {\color{red}\textbullet}) represent the net change in charge of a vertex. (b) The effect of an externally applied magnetic field on a vertical and sloped magnet for different values of the energy barrier. $E_\mathrm{MS}$ is the same for each $E_\mathrm{B}$, however it is higher for the sloped magnets compared to the vertical ones. When $E_\mathrm{B}<E_\mathrm{B}^c$ ($E_\mathrm{B}>E_\mathrm{B}^c$), the sloped (vertical) magnet will flip first since the Zeeman energy $E_\mathrm{Z}$ increases faster for the vertical magnets.}
    \label{fig:kagome_reversal}
\end{figure*}
In the experiment, the two reversal steps occurred gradually over a range of fields, resembling more of an S-curve rather than a sharp step. This is due to a random spread on the coercive field of each nanomagnet due to imperfections in lithography.~\cite{fraleigh2017characterization} We modeled this in the simulation by introducing a random Gaussian variation on the energy barrier, with $\sigma(E_\mathrm{B})/\langle E_\mathrm{B} \rangle=\qty{7}{\percent}$ yielding the closest agreement to the experiment. Note that the experimental hysteresis curve is not symmetric, in contrast to our simulations, which Li \textit{et al.} attributed to a small sample movement during their measurement which changed the applied field angle with respect to the array. \par 
As can be noted from the figure, the resulting hysteresis, and particularly the field magnitude for the second reversal step, is highly dependent on the choice of $\rho=\mu_\perp/\mu_\parallel$. When the asymmetry of the energy barrier is neglected ($\rho=0$), the field magnitude required for the second step is significantly overestimated, as seen in the blue curve of~\crefSubFigRef{fig:test_pinwheelReversal}{a} ($\approx\qty{60}{\milli\tesla}$ vs. experiment $\approx\qty{25}{\milli\tesla}$). Increasing $\rho$ reduces the field magnitude at which the second step occurs but has little effect on the first. Notably, for $\rho \gtrapprox 0.6$, the second step disappears entirely, and the reversal instead occurs by domain nucleation at the edges of the ASI, as shown in \crefSubFigRef{fig:test_pinwheelReversal}{e}. \par
Thus, $\rho \approx 0.4$ yields the best correspondence to the experiment. A possible explanation for this is that the switching of magnets in the experimental ASI does not occur by uniform rotation, as assumed by the N\'eel-Arrhenius switching law (\cref{eq:Néel}). Indications of non-uniform rotation or domain wall-mediated reversal were previously noticed by Morley~\textit{et al.}~\cite{VogelFulcherTammannFreezing}

\subsection{Kagome reversal} 
The in-plane kagome ASI (\crefSubFigRef{fig:ASIs}{g}) is a typical example of a frustrated system: every vertex is formed by three magnets that must obey the two-in/one-out or two-out/one-in ice rule. Each magnet has four NN located near its endpoints, making this lattice an interesting testing ground for the dumbbell model. Both the point dipole model~\cite{Chern2011} and dumbbell model~\cite{Moller2009} have been used in previous studies of kagome ASI. The point dipole model tends to significantly underestimate the NN interaction, while the dumbbell model increases the magnetostatic interaction energy between closely spaced magnets, potentially affecting the dynamics of the system. We therefore expect the dumbbell model to provide a more accurate simulation of the kagome lattice than the point dipole model.~\cite{flatspin,mengotti2011kagome} \par
To illustrate this, we used \hotspice{} to reproduce the reversal process of kagome ASI as observed by Mengotti~\textit{et al.}~\cite{mengotti2011kagome} They studied the hysteresis loop using XMCD to image the evolution of the system's microstate during a gradual increase of the external field near the coercive field. We simulated a kagome ASI consisting of 173 magnets whose energy barrier $E_\mathrm{B} = \qty{120}{\electronvolt}$ with a standard deviation $\sigma(E_\mathrm{B})/\langle E_\mathrm{B} \rangle=\qty{5}{\percent}$. An increasing magnetic field $\vb{B}_\mathrm{ext}$ was applied at an angle of \ang{-93.6} relative to the horizontal axis. This corresponds to a reverse field in the negative y-direction with a slight \ang{-3.6} offset to break symmetry between sloped magnets, resulting in the creation of ``Dirac strings'' of flipped magnets. \par
As the field strength increases, the Zeeman energy of the magnets rises, but this increase is faster for the vertical magnets than for the sloped ones. However, the magnetostatic energy is lowest for the vertical magnets in the initial state. The balance between these contributions results in a critical energy barrier $E_\mathrm{B}^\mathrm{c}$: when the energy barrier $E_\mathrm{B}$ is lower (higher) than this value, a sloped (vertical) magnet will flip first, as shown in~\crefSubFigRef{fig:kagome_reversal}{b}. Therefore, the dynamics may depend on the simulation method, as the dumbbell model increases the magnetostatic interaction energy between nearest neighbors. \par
The results, shown in~\crefSubFigRef{fig:kagome_reversal}{a}, reveal a qualitative difference between the dumbbell and dipole models. In the dumbbell model, sloped magnets on the boundaries of the ASI flip first, while for the dipole method vertical magnets in the bulk flip first.~\cite{DDG_Masterproef} In the experiment of Mengotti~\textit{et al.}~\cite{mengotti2011kagome}, Dirac strings were observed to originate from sloped magnets, which is in agreement with the dumbbell simulation. 
Additionally, researchers using the `flatspin' ASI simulator, which employs a dipole-based model, observed Dirac strings to originate from vertical magnets, similar to our results with a dipole model, highlighting the dumbbell model's ability to capture distinct dynamics.~\cite{flatspin}

\subsection{Reservoir computing by ``clocking''}
\label{sec:clocking}
\hotspice{} was developed to explore the potential of reservoir computing (RC) in ASI. RC is a machine learning framework where an input signal is applied to a nonlinear dynamical system, known as the ``reservoir''.~\cite{RC_unification,RC_RecentAdvances} The reservoir generates a high-dimensional nonlinear response, facilitating linear separation without the need to train the reservoir itself.~\cite{RC_ASI,maass_LSM} The essential characteristics of reservoirs -- short-term memory and high dimensionality -- are inherent to many physical systems, including ASI, making them suitable for direct use as a reservoir.~\cite{RC_RecentAdvances} This capability has already been demonstrated numerically for various ASI lattices.~\cite{RC_ASI,ASI_computation,hon2021numerical,Paper_PMA_ASI} To optimize the RC capability of ASI, it can be advantageous to manipulate the system in small, discrete steps without triggering a global magnetization reversal or ``avalanche'', a process that can be achieved through ``clocking'' protocols.~\cite{clocking-protocol}

\subsubsection{Clocking in pinwheel ASI}
A clocked input encoding scheme for pinwheel ASI has been proposed by Jensen \textit{et al.}~\cite{clocking-protocol} They applied an external magnetic field in a well-chosen direction to selectively affect one of the two pinwheel sublattices at a time. The field magnitude was carefully tuned such that only the magnets with the lowest effective energy barrier $\widetilde{E_\mathrm{B}}$ -- typically those at the boundary of a superferromagnetic domain -- would switch. The field direction was then changed to affect the other sublattice. This two-step clocking scheme allows domain wall boundaries to advance one step at a time, thereby enabling intermediate magnetic states. \par
Because Hotspice does not use the Stoner-Wohlfarth model as Jensen \textit{et al.}~\cite{clocking-protocol} did, we adjusted several system parameters to achieve clocking.~\cite{IL_Masterproef} NN interactions, and therefore inter-sublattice interactions, were enhanced by rotating all magnets in the pinwheel lattice anticlockwise by an additional angle $\alpha=\ang{4}$, resulting in two sublattices rotated by \ang{49} and \ang{-41} with respect to the array edges.~\cite{IL_Masterproef} As discussed in \cref{sec:Verification_IP_SquarePinwheel}, $\alpha$ should remain within a certain range to avoid the transition to IP square ASI with antiferromagnetic domains.~\cite{ApparentFMpinwheel} \par
\begin{figure*} 
    \centering
    \includegraphics[width=\textwidth]{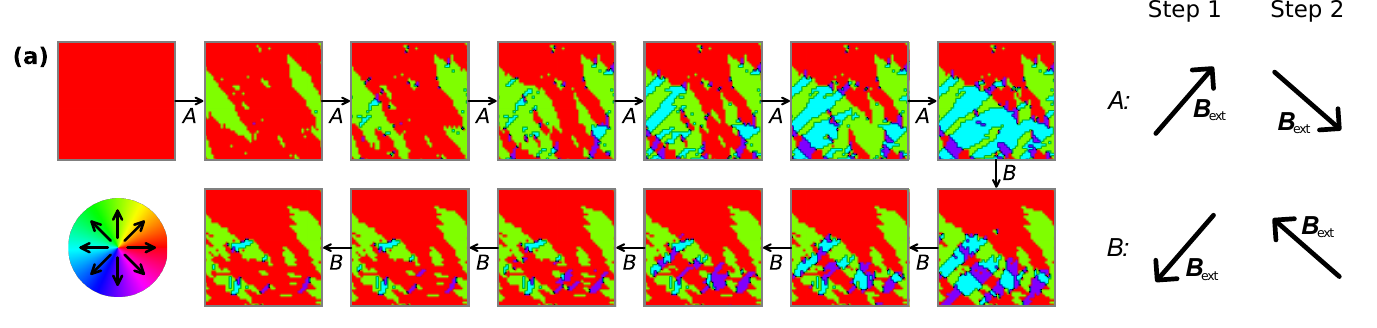}
    \includegraphics[width=\textwidth]{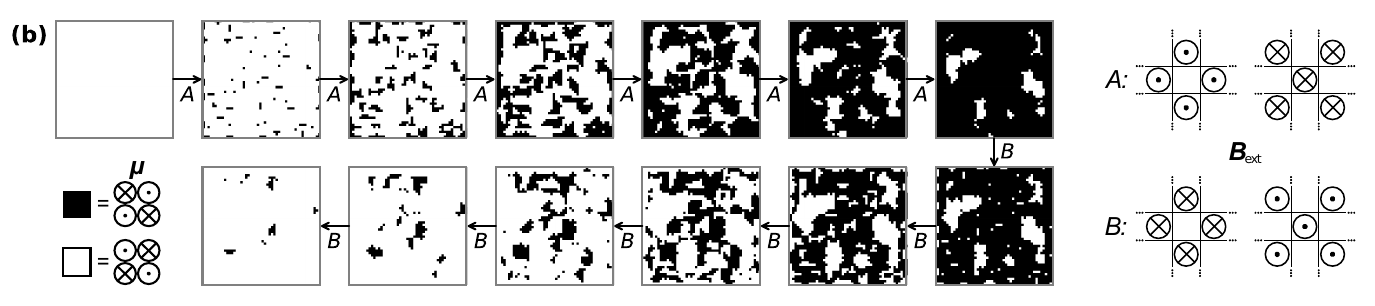}
    \caption{Effect of clocking schemes on the states of two ASI lattices. Six clocking cycles $A$ are followed by six clocking cycles $B$. These cycles are defined as shown on the right, and are different for the two lattices.
    (a) The clocking behavior of a $101 \times 101$ `diamond'-edge pinwheel ASI (containing 5100 magnets), with a lattice spacing $a = \qty{248}{\nm}$. Each magnet was rotated by $\alpha = \ang{4}$, has a magnetic moment $\mu = \qty{3e-16}{\ampere\meter\squared}$, and an energy barrier $E_\mathrm{B} = \qty{110}{\eV}$ with $\sigma(E_\mathrm{B})/\langle E_\mathrm{B} \rangle=\qty{5}{\percent}$ standard deviation. The asymmetric energy barrier was accounted for with $\rho=0.4$.
    All magnets are initially magnetized to the right (red).
    (b) A similar simulation for a $50 \times 50$ OOP square ASI, with circular magnets of radius \qty{85}{\nano\meter} and a lattice spacing $a = \qty{200}{\nano\meter}$. Each magnet has a magnetic moment $\mu = \qty{1.6e-16}{\ampere\meter\squared}$ and energy barrier $E_\mathrm{B}=\qty{60}{\electronvolt} \pm \qty{5}{\percent}$, similar to the OOP magnets used in \ccite{Paper_PMA_ASI}.
    All magnets are initially magnetized in one of the antiferromagnetic ground states (white). The magnetic field used in the clocking cycles has a magnitude $B_\mathrm{ext} = \qty{48}{\milli\tesla}$, and each step is applied for \qty{0.5}{\second}.
    }
    \label{fig:clocking}
\end{figure*}

We performed a simple test to visualize the behavior of pinwheel ASI under clocking, as shown in~\crefSubFigRef{fig:clocking}{a}. All magnets were initially magnetized to the right (red). Clocking cycle $A$ is defined by applying a magnetic field at $\ang{49}$ for $\qty{0.5}{\second}$, then at $\ang{-41}$ for another $\qty{0.5}{\second}$. This differs from the $\pm \ang{22}$ fields used by Jensen \textit{et al.}~\cite{clocking-protocol} Since switching occurs on nanosecond timescales, this duration is effectively infinite. After applying clocking cycle $A$ six times, the reverse cycle $B$ was applied six times as well. \par
Each time cycle $A$ or $B$ is applied, a well-defined effect is visible, though the impact of subsequent $B$ cycles diminishes as the system approaches its initial state. Note that, even though cycle $B$ applies the opposite fields to cycle $A$, it is not the inverse: the system does not return to the same states previously visited by cycles $A$. These results are qualitatively comparable to both experimental and numerical results achieved by Jensen \textit{et al.}~\cite{clocking-protocol} for ``AB clocking'', though each cycle in Hotspice switches magnets in a larger area than was observed by Jensen \textit{et al.}~\cite{clocking-protocol} Because of this nonlinear behavior which stores information in the state of the system by avoiding avalanches, this setup has the potential to perform well for reservoir computing purposes.~\cite{IL_Masterproef} \par
In summary, two key factors are necessary for controlled domain wall movement. First, the degeneracy between the ground states must be lifted, which is straightforward in pinwheel ASI as the four types of superferromagnetic domains respond differently to an in-plane field. Second, two independently addressable sublattices should exist to prevent avalanches; in pinwheel ASI, magnets of the two sublattices are perpendicular to each other, enabling their selective manipulation via in-plane fields.

\subsubsection{Clocking in OOP square ASI}
The concept of clocking can be extended to other ASI lattices that form domains, provided they contain multiple sublattices that can be addressed separately to avoid avalanches. As an example, we devised a clocking protocol for OOP square ASI based on the lessons learnt from pinwheel ASI. \par
In this system, neighboring magnets prefer anti-parallel alignment, resulting in two degenerate checkerboard ground states. One can therefore consider two sublattices -- akin to the black and white squares on a chess board. A uniform external field cannot move the domain walls because the domains have net zero magnetization. Instead, by applying opposite magnetic fields to each sublattice, domains of one type can be made to grow while the other shrinks. However, clocking also requires that only one sublattice is affected at a time. This can be achieved by first applying the field to one sublattice, removing it, and then applying an opposite field to the other sublattice. This procedure causes the domain walls to shift by up to two magnets per cycle, as demonstrated in~\crefSubFigRef{fig:clocking}{b}, where the gradual expansion of the desired domain type is clearly visible. \par
Implementing this procedure requires the use of local fields, which is more challenging than the global fields used for clocking pinwheel ASI. One potential solution is to use spin-orbit torque to flip the magnets, using current lines placed diagonally across the ASI to selectively affect one sublattice at a time. This clocking scheme benefits from disorder in the system (e.g., $\sigma(E_\mathrm{B}) > 0$ or vacancies), as this provides nucleation sites within the bulk.

\section{Conclusion}
We have presented in detail the underlying models used by \hotspice{}, an open-source and easily extensible Monte Carlo simulator designed to simulate ASI dynamics. In particular, we focused on the evaluation of several model variants extending beyond the basic Ising model typically used for ASI simulations. \par
Firstly, we considered more accurate calculations for magnetostatic interactions: OOP systems were found to benefit from a second-order correction, while IP systems achieve far more accurate results with a dumbbell model. Secondly, accounting for asymmetric switching channels proved key to e.g., correctly reproduce coercive fields in pinwheel ASI, provided that the reduced magnetization during non-coherent reversal processes was also accounted for. Finally, we provided two algorithms for handling switching events. The N\'eel-Arrhenius approach is best suited for simulating the temporal evolution of the system, including out-of-equilibrium dynamics. The Metropolis-Hastings algorithm efficiently explores equilibrium configurations, especially when multiple magnets switch simultaneously. These methods were compared to experimental and theoretical results for pinwheel, kagome, and square ASI. \par
This approach makes \hotspice{} complementary to traditional micromagnetic simulations. By sacrificing the detailed simulation of the internal magnetization structure of individual nanomagnets, the higher-level approximations employed by \hotspice{} enable the study of complex ASI dynamics in much larger systems and over significantly longer timescales. This opens new opportunities to use ASI for applied machine learning tasks like reservoir computing (RC). The ability of \hotspice{} to quickly sweep parameters and evaluate RC metrics facilitates the optimization of ASI configurations and the identification of suitable input protocols, as demonstrated by our simulation of clocking protocols in pinwheel and OOP square ASI. \par
In conclusion, \hotspice{}'s combination of speed, flexibility, and accuracy makes it a powerful tool to explore the rich physics of ASI systems and advance their use in innovative applications like reservoir computing.

\begin{acknowledgments}
The authors acknowledge funding from the \spinengine{} EU FET-Open Horizon 2020 RIA project (No. 861618) and the SHAPEme project (EOS ID 400077525) from the FWO and F.R.S.-FNRS under the Excellence of Science (EOS) program. J. L. is supported by the Research Foundation – Flanders (FWO) through senior postdoctoral research fellowship No. 12W7622N.
\end{acknowledgments}

\section*{Code availability}
\hotspice{} is open-source and available on \href{https://github.com/bvwaeyen/Hotspice}{GitHub}.


\bibliography{bibliography} 

\clearpage
\onecolumngrid
\pagenumbering{arabic}
\renewcommand{\thepage}{S\arabic{page}}
\renewcommand{\thesection}{S\arabic{section}}
\renewcommand{\thetable}{S\arabic{table}}
\renewcommand{\thefigure}{S\arabic{figure}}
\section*{Supplementary information}

The code used for the simulations presented in the main text is provided below. These are cleaned versions without plotting code, and not all \python{run()} functions return all the data shown in the figure. The full scripts, including plotting functions, can be found in the examples directory on the \hotspice{} \href{https://github.com/bvwaeyen/Hotspice}{GitHub} repository.

\subsection{Exchange-coupled OOP square system}
The following code was used for the simulation shown in Fig. 3a and 3b of the main text. It calculates the temperature-dependence of the average magnetization $\langle M \rangle/M_0 = \sum_i s_i/N$ and the nearest-neighbor correlation $\langle S_i S_{i+1} \rangle$ in an exchange-coupled OOP square ASI. Due to the absence of magnetostatic interactions, this simulation can significantly benefit from multi-switching using the Metropolis algorithm. Therefore, the simulation is performed on the GPU by setting the environment variable \python{"HOTSPICE_USE_GPU"} to \python{"True"}.
\begin{lstlisting}
import numpy as np

import os
os.environ["HOTSPICE_USE_GPU"] = "True" # We use Metropolis for multi-sampling on GPU
import hotspice

def run(T_range=np.linspace(0.9, 1.1, 21), N: int = 100, size: int = 800):
    """ `T_range` specifies the examined temperature range in multiples of T_c.
        At each step, `N` Monte Carlo steps per site are performed.
    """
    a = 1 # Choose large spacing to get many simultaneous Metropolis switches
    T_c = 1 # T_c = 2*J/(k_B*np.log(1+np.sqrt(2))), or the other way around:
    J = T_c*hotspice.kB*np.log(1 + np.sqrt(2))/2
    
    N_saved_iters = int(np.ceil(N/2)) # Half of N iters gets saved
    observable_shape = (T_range.size, N_saved_iters)
    m_avg = np.empty(observable_shape)
    NNcorr_avg = np.empty(observable_shape)
    
    mm = hotspice.ASI.OOP_Square(a, size, energies=[hotspice.ExchangeEnergy(J=J)], 
                                 pattern='uniform', PBC=True, T=1)
    mm.params.UPDATE_SCHEME = hotspice.Scheme.METROPOLIS
    for i, T in enumerate(T_range):
        mm.T = T*T_c
        mm.progress(t_max=np.inf, MCsteps_max=N/2) # Equilibrate for N/2 steps
        for j in range(N_saved_iters):
            mm.progress(t_max=np.inf, MCsteps_max=1)
            m_avg[i,j] = mm.m_avg
            NNcorr_avg[i,j] = mm.correlation_NN()
    return m_avg, NNcorr_avg

if __name__ == "__main__":
    size = 800 # With GPU+Metropolis, there is no reason to use size<800 here
    for N in [10, 100, 1600]:
        run(N=N, size=size)
\end{lstlisting}

\subsection{Exchange- and magnetostatically coupled OOP square system}
The following code was used for the simulation shown in Fig. 3c of the main text. It calculates the nearest-neighbor correlation $\langle S_i S_{i+1} \rangle$ in an OOP square ASI, where magnets are both magnetostatically and exchange-coupled. The effect of varying the relative strength of the exchange coupling with respect to the nearest-neighbor magnetostatic coupling: $\delta = E_{\mathrm{exch},i,j}/E_{\mathrm{MS},i,j} (j \in \mathcal{N}_i)$.
\begin{lstlisting}
import numpy as np

import os
os.environ["HOTSPICE_USE_GPU"] = "True" # We use Metropolis for multi-sampling on GPU
import hotspice

def run(delta_range=np.linspace(0, 4, 81), N: float = 100, size: int = 200):
    """ At each step, `N` Monte Carlo steps per site are performed. """
    T, a = 50, 1e-6 # A combination that is NOT in the paramagnetic state
    
    corr_avg, corr_std = np.empty_like(delta_range), np.empty_like(delta_range)
    states = np.empty(shape=(delta_range.size, size, size), dtype=bool)
    
    energyDD, energyExch = hotspice.DipolarEnergy(), hotspice.ExchangeEnergy()
    mm = hotspice.ASI.OOP_Square(a, size, E_B=0, T=T, energies=[energyDD, energyExch],
                                 pattern='AFM', PBC=True)
    mm.params.UPDATE_SCHEME = hotspice.Scheme.METROPOLIS
    for i, delta in enumerate(delta_range):
        energyExch.J = delta*energyDD.get_NN_interaction()/2 # Set delta by changing J
        corrs = [mm.correlation_NN()
                    for _ in mm.progress(t_max=np.inf, MCsteps_max=N, Q=np.inf)]
        half = len(corrs) // 2
        corr_avg[i], corr_std[i] = np.mean(corrs[half:]), np.std(corrs[half:])
        states[i,:,:] = (hotspice.utils.asnumpy(mm.m) + 1).astype(bool)
    return corr_avg, corr_std, states

if __name__ == "__main__":
    run()
\end{lstlisting}

\subsection{Non-interacting spin ensemble}
The following code was used for the simulation shown in Fig. 4a of the main text. It calculates the average magnetization $\langle M \rangle/M_0$ of a non-interacting ensemble of Ising spins when an external field of magnitude $B$ is applied.
\begin{lstlisting}
import numpy as np
import hotspice

def run(N: float = 10, size: int = 1000, monopoles: bool = False):
    """ Tests if the hyperbolic tangent is found when interactions are disabled.
        At each field magnitude, `N` Monte Carlo steps per spin are performed.
        An IP square ASI is used, with the field applied at 45 degrees to affect
        all magnets in the system equally.
    """
    T = 300
    moment = 1.1e-15
    alpha = np.linspace(0, 3, 11)

    energyZ = hotspice.energies.ZeemanEnergy(magnitude=0, angle=np.pi/4)
    mm = hotspice.ASI.IP_Square(1, size, moment=moment, energies=[energyZ],
                                pattern="random", T=T, m_perp_factor=0)
    mm.params.UPDATE_SCHEME = hotspice.Scheme.METROPOLIS

    mm.initialize_m('uniform', angle=np.pi/4)
    M_0 = mm.m_avg
    M_x = np.zeros_like(alpha)
    B_vals = alpha*hotspice.kB*T/moment*np.sqrt(2) # sqrt(2) due to 45deg field angle
    for i, B in enumerate(B_vals):
        energyZ.magnitude = B
        mm.progress(MCsteps_max=N)
        M_x[i] = mm.m_avg/M_0

if __name__ == "__main__":
    run()
\end{lstlisting}

\subsection{Square-to-pinwheel transition angle}
The following code was used for the simulation shown in Fig. 4b of the main text. It calculates the fraction of vertices with net zero magnetization, for several rotation angles of individual magnets between the square and pinwheel lattices.
\begin{lstlisting}
import numpy as np
import hotspice
from scipy import signal

def count_AFM_cells(mm: hotspice.Magnets):
    edges = np.logical_or(np.logical_or(mm.ixx == 0, mm.ixx == mm.nx - 1),
                          np.logical_or(mm.iyy == 0, mm.iyy == mm.ny - 1))
    vertices = np.where(np.logical_and(np.logical_not(edges), mm.ixx%2 == mm.iyy%2))
    mx = np.multiply(mm.m, mm.orientation[:,:,0])
    my = np.multiply(mm.m, mm.orientation[:,:,1])
    mask = np.array([[0, 1, 0], [1, 0, 1], [0, 1, 0]])
    boundary = "wrap" if mm.PBC else "fill"
    mx_avg = signal.convolve2d(mx, mask[::-1,::-1], mode="same", boundary=boundary)
    my_avg = signal.convolve2d(my, mask[::-1,::-1], mode="same", boundary=boundary)
    mx_valid = mx_avg[vertices]
    my_valid = my_avg[vertices]
    AFM = np.sum(np.logical_and(np.isclose(mx_valid, 0), np.isclose(my_valid, 0)))
    total = vertices[0].size
    return AFM/total

def run(N: float = 10, size: int = 51, monopoles: list[float] = None):
    """ At each angle, `N` Monte Carlo steps per spin are performed. """
    moment = 4.5e-17 # Paper value
    E_B = hotspice.utils.eV_to_J(15)
    init_pattern = "random"
    if monopoles is None: monopoles = [0]
    monopoles = np.asarray(monopoles).astype(float)
    
    angles = np.array([0, 5, 30, 32, 34, 36, 38, 40, 42, 44, 45])
    d_paper = 170e-9
    AFM_fraction = np.zeros((monopoles.size, angles.size, N))
    
    for k, monopole in enumerate(monopoles):
        monopole_d = 220e-9*monopole
        if monopole: energyDD = hotspice.energies.DiMonopolarEnergy(d=monopole_d)
        else: energyDD = hotspice.energies.DipolarEnergy()
        for j in range(N):
            for i, angle in enumerate(angles):
                a = d_paper*2*np.cos(np.deg2rad(angle))
                mm = hotspice.ASI.IP_Square(a, size, PBC=False, moment=moment, T=300, 
                                            energies=[energyDD], E_B=E_B,
                                            m_perp_factor=0.4, angle=np.deg2rad(angle))
                mm.params.UPDATE_SCHEME = hotspice.Scheme.METROPOLIS
                mm.initialize_m(init_pattern)
                mm.T = -np.min(mm.E)/hotspice.kB*0.3
                while mm.T_avg > 300:
                    mm.T *= 0.999
                    mm.update(Q=1)
                AFM_fraction[k,i,j] = count_AFM_cells(mm)

if __name__ == "__main__":
    run(monopoles=[0, 1], N=10)
\end{lstlisting}

\subsection{Pinwheel reversal}
The following code was used for the simulation shown in Fig. 5 of the main text. It performs a hysteresis on pinwheel ASI, for a range of the parameter $\rho=m_\perp/m_\parallel$, illustrating the importance of this parameter. The nanomagnet parameters were derived from the experimental setup in~\cite{li2018pinwheel}.
\begin{lstlisting}
import numpy as np
import hotspice

def run(size: int = 50, m_perp_factor: float = 1,
        monopoles: bool = False, angle: float = np.pi/6):
    """ A system size of 50 gives the exact same geometry as in the experiment, but
        with the x- and y-axes flipped. However, the experiment also defines the angle
        clockwise from the y-axis, which is therefore the same as our `B_angle`.
    """
    l, d, t, Msat = 470e-9, 170e-9, 10e-9, 800e3
    moment = Msat*((d*d*np.pi/4) + (l-d)*d)*t # [Am^2] moment of stadium-shaped magnet
    a = 420e-9*np.sqrt(2) # [m] Lattice spacing (for 420nm NN spacing in Pinwheel)
    T = 300 # [K] Assume room temperature
    E_B = hotspice.utils.eV_to_J(71) # [J] Energy barrier, calculated with mumax3
    E_B_std = 0.07
    
    if monopoles: energyDD = hotspice.energies.DiMonopolarEnergy(d=0.9*l, small_d=0.9*d)
    else: energyDD = hotspice.energies.DipolarEnergy()
    energyZ = hotspice.ZeemanEnergy()
    mm = hotspice.ASI.IP_Pinwheel(a, size, moment=moment, E_B=E_B, E_B_std=E_B_std,
                                  T=T, PBC=False, m_perp_factor=m_perp_factor,
                                  energies=[energyDD, energyZ])
    mm.params.UPDATE_SCHEME = hotspice.Scheme.NEEL
    
    B_max = 0.1 # [T] The hysteresis occurs between `-B_max` -> `B_max` -> `-B_max`.
    _N = 20000 # Number of `mm.update()` steps between `B_max` and `-B_max`.
    B_fields = np.linspace(-B_max, B_max, _N) # One sweep to opposite fields (N steps)
    B_fields = np.append(B_fields, np.flip(B_fields)) # Sweep both ways (2*N steps)
    m_avg, m_angle = np.zeros_like(B_fields), np.zeros_like(B_fields)
    
    mm.initialize_m(pattern='uniform', angle=angle+np.pi) # Initialize along field
    for i, B in enumerate(B_fields):
        energyZ.set_field(angle=angle, magnitude=B)
        mm.progress()
        m_avg[i], m_angle[i] = mm.m_avg, mm.m_avg_angle
    return B_fields, m_avg, m_angle

if __name__ == "__main__":
    for m_perp_factor in [0, 0.4, 1]: run(m_perp_factor=m_perp_factor)
\end{lstlisting}

\subsection{Kagome reversal}
The following code was used for the simulation shown in Fig. 6 of the main text. It performs a reversal of kagome ASI, using both the point dipole and dumbbell models, which affect the calculation of the magnetostatic energy. The nanomagnet parameters were derived from the experimental setup in~\cite{mengotti2011kagome}.
\begin{lstlisting}
import numpy as np
import hotspice

def run(size: int = 30, monopoles: bool = True, angle: float = np.pi/2 - 3.6*np.pi/180,
        E_B: float = hotspice.utils.eV_to_J(120), E_B_std: float = 0.05):
    a = 1e-6 # Gives 500nm NN center-to-center distance
    d = 470e-9  # Must be <577.35e-9, otherwise neighbors touch
    moment = 1.1e-15
    
    _N = 2000 # Number of `mm.update()` steps between `B_max` and `-B_max`.
    B_max = 0.06 # [T] The hysteresis occurs between `-B_max` -> `B_max` -> `-B_max`.
    B_fields = np.linspace(B_max, -B_max, _N) # One sweep to opposite fields (N steps)
    B_fields = np.append(B_fields, np.flip(B_fields)) # Sweep both ways (2*N steps)

    if monopoles: energyDD = hotspice.energies.DiMonopolarEnergy(d=d)
    else: energyDD = hotspice.energies.DipolarEnergy()
    energyZ = hotspice.ZeemanEnergy(angle=angle)
    mm = hotspice.ASI.IP_Kagome(a, size, moment=moment, energies=[energyDD, energyZ],
                                E_B=E_B, E_B_std=E_B_std, m_perp_factor=0)
    mm.params.UPDATE_SCHEME = hotspice.Scheme.NEEL
    
    # Perform the reversal
    mm.initialize_m("uniform", angle=angle)
    M_S = mm.m_avg_y
    M_y = np.zeros_like(B_fields)
    for i, B in enumerate(B_fields):
        energyZ.magnitude = B
        mm.progress()
        M_y[i] = mm.m_avg_y/M_S

    # Identify the critical field magnitude
    for i, _ in enumerate(M_y): # Identify the critical field magnitude
        if (M_y[i] <= 0 < M_y[i + 1]) or (M_y[i] >= 0 > M_y[i + 1]):
            B_C = np.abs(B_fields[i] - (M_y[i]*(B_fields[i+1] - B_fields[i]))
                         /(M_y[i+1] - M_y[i]))
            break

    # Save states at certain field magnitudes
    states, fields = [], [-0.85, -0.92, -0.99, -1.06]
    for B in fields:
        energyZ.magnitude = B*B_C
        mm.progress()
        states.append(np.where(mm.occupation == 0, np.nan, mm.m))
    return states, fields

if __name__ == "__main__":
    run(monopoles=False)
    run(monopoles=True)
\end{lstlisting} 

\subsection{Pinwheel clocking}
The following code was used for the simulation shown in Fig. 7a of the main text. It demonstrates the use of the \python{hotspice.io} module to apply clocked input to pinwheel ASI by defining a custom \python{Datastream} and \python{Inputter}. These are then used to apply the clocking cycles $A$ and $B$ as described in the main text. The nanomagnet parameters were derived from the experimental setup in~\cite{clocking-protocol}.
\begin{lstlisting}
import numpy as np
import hotspice

class BinaryListDatastream(hotspice.io.BinaryDatastream):
    """ Returns values from a given list. """
    def __init__(self, binary_list: list[int] = [0]):
        self.binary_list = binary_list
        self.len_list = len(binary_list)  # will be used often, no need to recalculate
        self.index = 0  # start at the beginning
        super().__init__()

    def get_next(self, n=1) -> np.ndarray:
        """ Returns next `n` values as xp.ndarray. Loops back to start at the end. """
        values = [self.binary_list[(self.index + i) % self.len_list] for i in range(n)]
        self.index += n
        return np.array(values)

class ClockingFieldInputter(hotspice.io.FieldInputter):
    def __init__(self, datastream: hotspice.io.BinaryDatastream, magnitude=1,
                 angle=0, spread=np.pi/8, n=2, frequency=1):
        """ Applies an external field at `angle+spread` rad for 0.5/frequency seconds,
            then at `angle-spread` rad for bit 0. It does the same +180deg for bit 1.
            This avoids avalanches by only affecting one sublattice at a time.
        """
        self.spread = spread
        super().__init__(datastream, magnitude=magnitude,
                         angle=angle, n=n, frequency=frequency)

    def bit_to_angles(self, bit):
        if not bit: return (self.angle + self.spread, self.angle - self.spread)
        return (self.angle + self.spread + np.pi, self.angle - self.spread + np.pi)

    def input_single(self, mm: hotspice.ASI.IP_ASI, value: bool|int):
        if self.frequency and mm.params.UPDATE_SCHEME != hotspice.Scheme.NEEL:
            raise AttributeError("Can only use frequency if UPDATE_SCHEME is NEEL.")
        if not mm.in_plane:
            raise AttributeError("Can only use ClockingFieldInputter on in-plane ASI.")

        Zeeman: hotspice.ZeemanEnergy = mm.get_energy('Zeeman')
        if Zeeman is None: mm.add_energy(Zeeman := hotspice.ZeemanEnergy(0, 0))
        angle1, angle2 = self.bit_to_angles(value)
        Zeeman.set_field(magnitude=self.magnitude, angle=angle1)
        mm.progress(t_max=0.5/self.frequency, MCsteps_max=self.n)
        Zeeman.set_field(magnitude=self.magnitude, angle=angle2)
        mm.progress(t_max=0.5/self.frequency, MCsteps_max=self.n)

def run(N: int = 13, size: int = 101, E_B_std: float = 0.05,
        m_perp_factor: float = 0.4, magnitude: float = 53e-3):
    a = 248e-9 # Lattice spacing
    moment = 3e-16
    T = 1 # For deterministic switching
    E_B = hotspice.utils.eV_to_J(110)
    np.random.seed(2) # 2 gives a nice result.
    
    spread = np.pi/4 # We use 45deg instead of the 22.5deg used in the Clocking paper.
    bits = [None] + [1]*(N//2) + [0]*(N//2)
    datastream = BinaryListDatastream(bits)
    inputter = ClockingFieldInputter(datastream, magnitude=magnitude, spread=spread)
    lattice_angle = np.deg2rad(4) # Rotate all magnets by 4deg
    mm = hotspice.ASI.IP_Pinwheel(a, size, moment=moment, E_B=E_B, E_B_std=E_B_std,
                                  T=T, pattern="uniform", angle=lattice_angle,
                                  m_perp_factor=m_perp_factor)
    mm.params.UPDATE_SCHEME = hotspice.Scheme.NEEL # NEEL is best at low T and high E_B
    mm.add_energy(hotspice.ZeemanEnergy())

    states = []
    for bit in datastream.binary_list:
        if bit is not None: inputter.input(mm, values=[bit])
        states.append(np.copy(mm.m))
    return states

if __name__ == "__main__":
    run(m_perp_factor=0.4)
\end{lstlisting}

\subsection{OOP square clocking}
The following code was used for the simulation shown in Fig. 7b of the main text. It demonstrates a clocking protocol for OOP square ASI, available in the standard \hotspice{} distribution. The code applies the clocking cycles $A$ and $B$ as described in the main text.
\begin{lstlisting}
import numpy as np
import hotspice

def run(N: int = 13, magnitude: float = 0.05, E_B_std: float = 0.05,
        E_B: float = hotspice.utils.eV_to_J(60), moment: float = 2.37e-16,
        a: float = 230e-9, d: float = 0, vacancy_fraction: float = 0., size: int = 50):
    mm = hotspice.ASI.OOP_Square(a, size, E_B=E_B, E_B_std=E_B_std,
                                 moment=moment, major_axis=d, minor_axis=d)
    mm.params.UPDATE_SCHEME = hotspice.Scheme.NEEL
    mm.add_energy(hotspice.ZeemanEnergy())
    vacancies = int(mm.n*vacancy_fraction)
    mm.occupation[np.random.randint(mm.ny, size=vacancies),
                  np.random.randint(mm.nx, size=vacancies)] = 0
    ds = hotspice.io.RandomBinaryDatastream()
    inputter = hotspice.io.OOPSquareChessStepsInputter(ds, magnitude=magnitude)
    
    states, domains = [], []
    values = []
    mm.initialize_m('AFM', angle=np.pi)
    for i in range(N):
        if i == 0: values.append(None)
        else: values.append(inputter.input(mm, values=(i < (N//2 + 1)))[0])
        states.append(np.where(mm.occupation == 0, np.nan, mm.m))
        domains.append(np.where(mm.occupation == 0, np.nan, mm.get_domains()))
    return states, domains

if __name__ == "__main__":
    run(magnitude=0.048, a=200e-9, d=170e-9)
\end{lstlisting}


\end{document}